\documentstyle[preprint,eqsecnum,aps]{revtex}
\begin{document}

%my macros for LaTeX fixes

% special math symbols:
\def\Rnum{{\bf R}}
\def\Cnum{{\bf C}}

% refs: 
%put `( )' around an equation ref in latex
\def\eqref#1{(\ref{#1})}
\def\eqrefs#1#2{(\ref{#1}) and~(\ref{#2})}
\def\eqsref#1#2{(\ref{#1}) to~(\ref{#2})}

%put `Eq.( )' around an equation ref in latex
\def\Eqref#1{Eq.~(\ref{#1})}
\def\Eqrefs#1#2{Eqs.~(\ref{#1}) and~(\ref{#2})}
\def\Eqsref#1#2{Eqs.~(\ref{#1}) to~(\ref{#2})}

%put `Sec. ' before a section ref in latex
\def\secref#1{Sec.~\ref{#1}}
\def\secrefs#1#2{Secs.~\ref{#1} and~\ref{#2}}
\def\secsref#1#2{Secs.~\ref{#1} to~\ref{#2}}

%put `App. ' before an appendix ref in latex
\def\appref#1{App.~\ref{#1}}

%put `Ref. ' before a bibitem ref in latex
\def\Ref#1{Ref.\cite{#1}}
\def\Refs#1{Refs.\cite{#1}}

%use footnote style for a bibitem ref in latex
\def\Cite#1{${\mathstrut}^{\cite{#1}}$}

%put `Table ' before a table ref in latex
\def\tableref#1{Table~\ref{#1}}

%put `Fig. ' before a figure ref in latex
\def\figref#1{Fig.~\ref{#1}}

% fix hyphenations to be 
\hyphenation{Eq Eqs Sec App Ref Fig}

% abbrevs for latex commands:
%equations
\def\EQ{\begin{equation}}
\def\EQs{\begin{eqnarray}}
\def\endEQ{\end{equation}}
\def\endEQs{\end{eqnarray}}

\def\eqtext#1{\hbox{\rm{#1}}}

% proclamations
\def\proclaim#1{\medbreak
\noindent{\it {#1}}\par\medbreak}
\def\Proclaim#1#2{\medbreak
\noindent{\bf {#1}}{\it {#2}}\par\medbreak}

%my macros

\def\fewquad{\qquad\qquad}
\def\severalquad{\qquad\fewquad}
\def\manyquad{\qquad\severalquad}
\def\manymanyquad{\manyquad\manyquad}

\def\sp#1{\vskip #1pt}

\def\sub#1{
\setbox1=\hbox{{$\scriptscriptstyle #1$}} 
\dimen1=0.6\ht1
\mkern-2mu \lower\dimen1\box1 \hbox to\dimen1{\box1\hfill} }

\def\eqtext#1{\hbox{\rm{#1}}}

\def\endproof{
\setbox2=\hbox{{$\sqcup$}} \setbox1=\hbox{{$\sqcap$}} 
\dimen1=\wd1
\box2\kern-\dimen1 \hbox to\dimen1{\box1} }

\def\mstrut{\mathstrut}
\def\hp#1{\hphantom{#1}}

\def\mixedindices#1#2{{\mstrut}^{\mstrut #1}_{\mstrut #2}}
\def\downindex#1{{\mstrut}^{\mstrut}_{\mstrut #1}}
\def\upindex#1{{\mstrut}_{\mstrut}^{\mstrut #1}}
\def\downupindices#1#2{{\mstrut}_{\mstrut #1}^{\hp{#1}\mstrut #2}}
\def\updownindices#1#2{{\mstrut}^{\mstrut #1}_{\hp{#1}\mstrut #2}}

\def\index#1{{\scriptstyle #1}}

\def\Parder#1#2{
\mathchoice{\partial{#1} \over\partial{#2}}{\partial{#1}/\partial{#2}}{}{} }
\def\parder#1{\partial/\partial{#1}}

\def\L#1{L\downindex{#1}}
\def\E{{\cal E}}
\def\J#1{J\downindex{#1}}
\def\Q#1{Q\downindex{#1}}
\def\C{C}
\def\B#1{{\tilde B}\downindex{#1}}
\def\T#1{\Theta\downindex{#1}}
\def\w#1{\omega\downindex{#1}}
\def\btT#1{\psi\downindex{#1}}
\def\dyn{{}_{\cal D}}
\def\nondyn{{}_{\cal N}}
\def\bc{{\cal F}}
\def\H#1{H\upindex{\rm #1}}
\def\bcdata#1#2{{\cal F}\mixedindices{#2}{#1}}
\def\othbcdata#1#2{\hat{\cal F}\mixedindices{#2}{#1}}
\def\HE{{\cal E}_H}
\def\Hdens{{\cal H}}

\def\region{{\cal V}}
\def\bcsurface{{\cal B}}

\def\tfvec#1{\xi\upindex{#1}}
\def\tfduvec#1{\xi\downindex{#1}}
\def\perptfvec#1{\zeta\upindex{#1}}
\def\tantfvec#1{\xi\mixedindices{#1}{\parallel}}

\def\der#1{\partial\downindex{#1}}
\def\coder#1{\partial\upindex{#1}}
\def\nder#1#2{\partial\mixedindices{#1}{#2}}
\def\D#1{{\cal D}\downindex{#1}}
\def\coD#1{{\cal D}\upindex{#1}}
\def\covder#1{\nabla\downindex{#1}}
\def\covcoder#1{\nabla\upindex{#1}}
\def\covSder#1{\nabla\mixedindices{S}{#1}}
\def\covcoSder#1{\nabla\upindex{S #1}}
\def\perpD#1{{\cal D}\mixedindices{\perp}{#1}}
\def\perpcoD#1{{\cal D}\upindex{\perp #1}}
\def\perpcovder#1{\nabla\mixedindices{\perp}{#1}}

\def\covgder#1{{{}^g\nabla}\downindex{#1}}
\def\covgcoder#1{{{}^g\nabla}\upindex{#1}}
\def\div#1{\partial\mixedindices{#1}{\Sigma}}
\def\codiv#1{\partial\mixedindices{\Sigma}{#1}}

\def\g#1#2{g\downupindices{#1}{#2}}
\def\metric#1#2{\sigma\downupindices{#1}{#2}}
\def\perpmetric#1#2{\sigma\mixedindices{\perp}{#1}\upindex{#2}}
\def\flat#1#2{\eta\downupindices{#1}{#2}}
\def\vol#1#2{\epsilon\downupindices{#1}{#2}}
\def\duvol#1#2{{*\epsilon}\downupindices{#1}{#2}}
\def\perpvol#1#2{\epsilon\mixedindices{\perp}{#1}\upindex{#2}}
\def\coordvol#1#2{\varepsilon\downupindices{#1}{#2}}
\def\e#1#2{e\downupindices{#1}{#2}}
\def\inve#1#2{e\updownindices{#1}{#2}}
\def\x#1#2{x\mixedindices{#1}{#2}}
\def\id#1#2{\delta\mixedindices{#2}{#1}}

\def\curv#1#2{R\downupindices{#1}{#2}}
\def\scurv{R}
\def\con#1#2{\Gamma\updownindices{#1}{#2}}
\def\tcurv#1#2{{\tilde R}\downupindices{#1}{#2}}
\def\curvS#1#2{{\cal R}\downupindices{#1}{#2}}
\def\scurvS{{\cal R}}
\def\excurv#1#2{\kappa\mixedindices{#1}{#2}}
\def\trexcurv{\kappa}
\def\norScon#1#2{{\cal J}\mixedindices{#1\perp}{#2}}
\def\norScurv#1#2{{\cal R}\mixedindices{\perp}{#1}\upindex{#2}}
\def\perpcon#1#2{J\mixedindices{#1\perp}{#2}}
\def\perpcurv#1#2{R\mixedindices{\perp}{#1}\upindex{#2}}

\def\surfcurv#1#2{{\cal R}\downupindices{#1}{#2}}
\def\surfexcurv#1#2{{\cal K}\downupindices{#1}{#2}}
\def\surfder#1#2{{\cal D}\mixedindices{#2}{#1}}
\def\surfaccel#1{a\downindex{#1}}

\def\EMT#1#2{T\downupindices{#1}{#2}}
\def\G#1#2{G\downupindices{#1}{#2}}

\def\a#1#2{\alpha\mixedindices{#1}{#2}}
\def\ta#1#2{\tilde\alpha\mixedindices{#1}{#2}}
\def\b#1#2{\beta\mixedindices{#1}{#2}}
\def\tb#1#2{\tilde\beta\mixedindices{#1}{#2}}
\def\asub#1#2#3{\alpha_{#1}\mixedindices{#2}{#3}}

\def\s#1#2{s\mixedindices{#1}{#2}}
\def\t#1#2{t\mixedindices{#1}{#2}}
\def\h#1#2{h\downupindices{#1}{#2}}
\def\K#1#2{K\downupindices{#1}{#2}}
\def\trK{\K{}{}}
\def\N#1#2{N\mixedindices{#1}{#2}}
\def\tN#1#2{\tilde N\mixedindices{#1}{#2}}

\def\n#1#2{n\mixedindices{#1}{#2}}
\def\q#1#2{q\downupindices{#1}{#2}}
\def\p#1#2{p\downupindices{#1}{#2}}
\def\u#1#2{u\mixedindices{#1}{#2}}

\def\svec{\vec s}

\def\news#1#2{s\mixedindices{#1}{#2}}
\def\newt#1#2{t\mixedindices{#1}{#2}}
\def\newh#1#2{h\downupindices{#1}{#2}}
\def\newK#1#2{K\downupindices{#1}{#2}}
\def\newtrK{\K{}{}}

\def\coeff#1#2#3{\Pi_{#1}\mixedindices{#3}{#2}}
\def\othcoeff#1#2#3{\hat\Pi_{#1}\mixedindices{#3}{#2}}

\def\frame#1#2{\theta\mixedindices{#1}{#2}}
\def\outnframe#1#2{\theta\mixedindices{+#1}{#2}}
\def\innframe#1#2{\theta\mixedindices{-#1}{#2}}
\def\nframe#1#2{\theta\mixedindices{#1}{#2}}
\def\fixinframe#1#2{\hat\theta{}\mixedindices{+#1}{#2}}
\def\fixoutframe#1#2{\hat\theta{}\mixedindices{-#1}{#2}}

\def\frcon#1#2{\Gamma\downupindices{#1}{#2}}
\def\tfrcon#1#2{{\tilde \Gamma}\downupindices{#1}{#2}}
\def\adfr#1#2{\vartheta\mixedindices{#1}{#2}}
\def\adothfr#1#2{\tilde\vartheta\mixedindices{#1}{#2}}

\def\U#1#2{U\updownindices{#1}{#2}}

\def\mcurvH#1#2{H\mixedindices{#1}{#2}}
\def\perpmcurvH#1#2{H\mixedindices{#1}{\perp #2}}
\def\Pperp#1#2#3{(P^{#1}_\perp)\mixedindices{#2}{#3}}
\def\Ppar#1#2#3{(P^{#1}_\parallel)\mixedindices{#2}{#3}}
\def\normH{|\H{}{}|}
\def\normPperp{|P_\perp|}
\def\hatH#1#2{\hat H\mixedindices{#1}{#2}}
\def\hatperpH#1#2{\hat H\mixedindices{#1}{\perp #2}}
\def\hatP#1#2{\hat \zeta\mixedindices{#1}{#2}}
\def\fixPpar#1#2{({\hat P}_\parallel)\mixedindices{#1}{#2}}
\def\fixPperp#1#2{({\hat P}_\perp)\mixedindices{#1}{#2}}
\def\fixP#1#2{{\hat P}\mixedindices{#1}{#2}}
\def\P#1#2{P\mixedindices{#1}{#2}}
\def\barP#1#2{\bar P\mixedindices{#1}{#2}}

\def\bs#1#2{s'\mixedindices{#1}{#2}}
\def\bt#1#2{t'\mixedindices{#1}{#2}}
\def\ha{\chi}

\def\fixs#1#2{\hat s\mixedindices{#1}{#2}}
\def\fixt#1#2{\hat t\mixedindices{#1}{#2}}

\def\fixu#1#2{\hat u\mixedindices{#1}{#2}}
\def\fixv#1#2{\hat v\mixedindices{#1}{#2}}

\def\v#1#2{v\mixedindices{#1}{#2}}

\def\PrS{{\cal P}_S}
\def\PrperpS{{\cal P}_S^\perp}
\def\PrSop#1#2{(\PrS)\downupindices{#1}{#2}}
\def\PrperpSop#1#2{(\PrperpS)\downupindices{#1}{#2}}
\def\Pr#1{{\cal P}_{#1}}
\def\Prop#1#2#3{({\cal P}_{#1})\downupindices{#2}{#3}}
\def\Prperp#1{{\cal P}^\perp_{#1}}
\def\Prperpop#1#2#3{({\cal P}^\perp_{#1})\downupindices{#2}{#3}}

\def\ncongruenceS{S_{(\lambda_+,\lambda_-)}}

\def\TS{T(S)}
\def\TperpS{T(S)^\perp}
\def\TM{T(M)}

\def\Lie#1{{\cal L}_{#1}}

\def\O{O}
\def\spi{\iota^0}
\def\scri{{{\cal I}^\pm}}

% Maxwell macros

\def\A#1{A\downindex{#1}}
\def\F#1{F\downindex{#1}}
\def\duF#1{{*F}\downindex{#1}}
\def\coF#1{F\upindex{#1}}

\def\Efield#1{E\downindex{#1}}
\def\Bfield#1{B\downindex{#1}}

\def\Evec#1{{\vec E}_{#1}}
\def\Bvec#1{{\vec B}_{#1}}
\def\Avec#1{{\vec A}_{#1}}
\def\dervec#1{\vec\partial\downindex{#1}}

% Yang-Mills macros

\def\ymA#1#2{A\mixedindices{\mit#1}{#2}}
\def\ymF#1#2{F\mixedindices{\mit#1}{#2}}
\def\ymcoF#1#2{F\upindex{{\mit#1}#2}}
\def\ymduF#1#2{{*F}\mixedindices{\mit#1}{#2}}
\def\C#1#2{C\downupindices{\mit#1}{\mit#2}}
\def\k#1#2{k\downupindices{\mit#1}{\mit#2}}
\def\duC#1#2{C^*\updownindices{\mit#1}{\mit#2}}

\def\ymconn#1#2#3{\Gamma\mixedindices{\mit#2}{#1}\downindex{\mit#3}}
\def\ymcoconn#1#2#3{\Gamma\upindex{#1{\mit#2}}\downindex{\mit#3}}

\def\W#1#2{W\mixedindices{\mit#1}{#2}}
\def\coW#1#2{W\upindex{{\mit#1}#2}}

\def\ymU#1#2{U\updownindices{\mit#1}{\mit#2}}
\def\yminvU#1#2{U\updownindices{-1{\mit#1}}{\mit#2}}

% Klein-Gordon Higgs macros

\def\kg#1{\varphi^{\mit#1}}

\def\ie/{i.e.}
\def\eg/{e.g.}
\def\cf/{c.f.}
\def\const{{\rm const}}

\hyphenation{
}

\def\onfr/{orthonormal frame}
\def\bdc/{boundary condition}
\def\bdt/{boundary term}
\def\af/{asymptotically flat}
\def\KG/{Klein-Gordon}
\def\YM/{Yang-Mills}

%end of macros

\draft

\title{Properties of the symplectic structure of General Relativity\\
for spatially bounded spacetime regions.}

\author{Stephen C. Anco
\thanks{Email address : sanco@brocku.ca}}

\address{Department of Mathematics, Brock University\\
St Catharines, Ontario, L2S 3A1, Canada}

\author{Roh S. Tung
\thanks{Email address : roh@gr.uchicago.edu}}

\address{Enrico Fermi Institute, University of Chicago\\
Chicago, Illinois 60637, USA}

%\date{\today}

\maketitle

\begin{abstract}
We continue a previous analysis of 
the covariant Hamiltonian symplectic structure 
of General Relativity for spatially bounded regions of spacetime. 
To allow for wide generality, 
the Hamiltonian is formulated using any fixed hypersurface, 
with a boundary given by a closed spacelike 2-surface. 
A main result is that we obtain Hamiltonians associated to 
Dirichlet and Neumann boundary conditions on the gravitational field
coupled to matter sources, in particular 
a Klein-Gordon field,
an electromagnetic field, 
and a set of Yang-Mills-Higgs fields. 
The Hamiltonians are given by a covariant form of 
the Arnowitt-Deser-Misner Hamiltonian 
modified by a surface integral term 
that depends on the particular boundary conditions. 
The general form of this surface integral involves 
an underlying ``energy-momentum'' vector in the spacetime tangent space
at the spatial boundary 2-surface. 
We give examples of the resulting Dirichlet and Neumann vectors 
for topologically spherical 2-surfaces in Minkowski spacetime, 
spherically symmetric spacetimes, 
and stationary axisymmetric spacetimes. 
Moreover, we establish the relation between these vectors 
and the ADM energy-momentum vector for a 2-surface taken in a limit
to be spatial infinity in asymptotically flat spacetimes. 
We also discuss the geometrical properties of 
the Dirichlet and Neumann vectors 
and obtain several striking results relating these vectors 
to the mean curvature and normal curvature connection of the 2-surface. 
Most significantly, 
the part of the Dirichlet vector normal to the 2-surface 
depends only on the spacetime metric at this surface
and thereby defines a geometrical normal vector field 
on the 2-surface. 
We show that this normal vector 
is orthogonal to the mean curvature vector, 
and its norm is the mean null extrinsic curvature, 
while its direction is such that there is zero expansion of the 2-surface
\ie/ the Lie derivative of the surface volume form in this direction
vanishes. 
This leads to a direct relation between the Dirichlet vector
and the condition for a spacelike 2-surface to be (marginally) trapped. 
\end{abstract}


%\narrowtext
\tightenlines

\section{Introduction}

In a previous paper \cite{Anco-TungI} 
we began an investigation of 
the covariant symplectic structure associated to 
the Einstein equations for the gravitational field 
in any fixed spatially compact region $\Sigma\times\Rnum$ of spacetime
whose spacelike slices $\Sigma$ possess 
a closed 2-surface boundary $\partial\Sigma$, 
with a fixed time-flow vector field 
tangent to the timelike boundary hypersurface $\partial\Sigma\times\Rnum$.
Through an analysis of \bdc/s
required for existence of a Hamiltonian variational principle,
we derived Dirichlet, Neumann, and mixed type boundary conditions
for the spacetime metric at the spatial boundary 2-surface $\partial\Sigma$.
The corresponding Hamiltonians we obtained are given by 
a covariant form of the ADM Hamiltonian
plus a surface integral term whose form depends on the \bdc/s. 
We also showed that these Hamiltonians naturally yield
covariant field equations which are equivalent to a 3+1 split of
the Einstein equations into the well-known 
constraint equations and geometrical time-evolution equations
for the spacetime metric. 

The present paper continues the previous analysis in two significant ways. 
First, in \secref{matter},
we investigate the covariant symplectic structure of 
the Einstein equations coupled to matter sources
in any fixed spatially bounded region of spacetime. 
Specifically, 
we consider Dirichlet and Neumann \bdc/s for 
a scalar field, an electromagnetic field, 
and a set of Yang-Mills/Higgs fields. 
Furthermore, 
we allow the fixed time-flow vector field on spacetime 
to have an arbitrary direction (\ie/ not necessarily timelike)
in the spacetime region. 
Such freedom in the choice of the time-flow vector field 
is useful for relating the Hamiltonian \bdt/s 
to expressions for total energy, momentum, angular momentum 
associated to the gravitational and matter fields 
on given hypersurfaces in spacetime. 

Next, in \secref{properties},
we discuss in detail the geometrical structure of
the gravitational part of the Dirichlet and Neumann Hamiltonian \bdt/s. 
In particular, as noted in \Ref{Anco-TungI}, 
these each involve an underlying locally constructed 
``energy-momentum'' vector at each point in the tangent space 
at the 2-surface. 
We show that the form of the \bdt/ vectors is closely related to 
the mean curvature vector and normal curvature connection 1-form
which describe the extrinsic geometry of the spatial boundary 2-surface
in spacetime. 
Most striking, 
we further show that the part of the Dirichlet \bdt/ vector 
orthogonal to the 2-surface 
yields a direction in which the 2-surface has zero expansion
in spacetime. 

Finally, through several examples, 
we illustrate the properties of 
the Dirichlet and Neumann \bdt/ vectors 
for topologically spherical 2-surfaces in various physically interesting
spacetimes in \secref{examples}. 
As a main result,
we show that in asymptotically flat spacetimes
the Dirichlet vector at spatial infinity 
can be identified in a natural way 
with the ADM energy-momentum vector. 

We make some concluding remarks in \secref{conclusion}. 
(The notation and conventions of \Ref{Anco-TungI} are used throughout.)

\section{ Matter fields }
\label{matter}

It is convenient here 
to employ the tetrad formulation of the Einstein equations,
since this simplifies the analysis of \bdc/s and Hamiltonian \bdt/s
as shown in \Ref{Anco-TungI}. 
We focus on Dirichlet and Neumann \bdc/s 
and make some remarks on more general \bdc/s at the end. 

\subsection{ Preliminaries }

On a given smooth orientable spacetime $(M,\g{ab}{})$, 
let $\tfvec{a}$ be a complete, smooth time-flow vector field, 
allowed to be timelike, spacelike, or null. 
Let $\Sigma$ be a region contained in a fixed hypersurface in $M$ 
such that the boundary of the region is 
a closed orientable spacelike 2-surface $\partial\Sigma$
(with the hypersurface allowed to be otherwise arbitrary). 

For treatment of \bdc/s when the time-flow $\tfvec{a}$ 
is not necessarily timelike, 
it is helpful to introduce the following structure associated to 
the boundary 2-surface $\partial\Sigma$.

Let $\Pr{\partial\Sigma}$ and $\Prperp{\partial\Sigma}$ 
denote projection operators onto the tangent subspaces 
$T(\partial\Sigma)$ and $T(\partial\Sigma)^\perp$ 
with respect to the surface $\partial\Sigma$ 
in local coordinates in $M$. 
Note $\Pr{\partial\Sigma}+\Prperp{\partial\Sigma}$ is the identity map 
on the tangent space $T(M)$ at $\partial\Sigma$. 
Define the metric on $\partial\Sigma$ by 
\EQ
\metric{}{ab} = \Pr{\partial\Sigma}( \g{}{ab} ) ,\quad
\metric{ab}{} = \g{ac}{} \g{bd}{} \metric{}{bc}
\endEQ
Let $\vol{ab}{}$
be the metric volume form on $\partial\Sigma$, 
and define 
\EQ\label{duvolid}
\duvol{ab}{} =\vol{}{cd} \vol{abcd}{}(g) 
\endEQ
in terms of the spacetime volume form $\vol{abcd}{}(g)$. 
Note that 
$\Pr{\partial\Sigma}( \duvol{ab}{} )
= \Prperp{\partial\Sigma}( \vol{ab}{} )
= 0$. 
A useful identity is given by 
\EQ\label{Prvolid}
2 \Prop{\partial\Sigma}{[a}{c} \Prop{\partial\Sigma}{b]}{d}
=\vol{ab}{} \vol{}{cd} . 
\endEQ

Let 
\EQ
\perptfvec{a} = \Prperp{\partial\Sigma}( \tfvec{a} ) ,\quad
\N{a}{} = \Pr{\partial\Sigma}( \tfvec{a} ) , 
\endEQ
so $\tfvec{a}= \perptfvec{a}+\N{a}{}$ 
decomposes into a sum of normal and tangential vectors
with respect to $\partial\Sigma$.
We now suppose $\tfvec{a}$ is not tangential to $\partial\Sigma$, 
\ie/ $\perptfvec{a}\neq 0$ everywhere on the surface $\partial\Sigma$.
In this situation, 
much of the formalism and results given in Sec.~3 of \Ref{Anco-TungI}
can be paralleled. 

Let $\bcsurface$ denote the hypersurface given by 
the image of $\partial\Sigma$ under a one-parameter diffeomorphism
generated by $\tfvec{a}$ on $M$. 
Note that the dual vector field $\duvol{ab}{} \tfvec{b}$
is hypersurface orthogonal 
since it is annihilated by all tangent vectors 
(in particular $\tfvec{a}$) in $\bcsurface$. 
Define a basis $\{ \news{}{a},\newt{}{a} \}$ 
for $T^*(\partial\Sigma)^\perp$
by diagonalization of the identity map
\EQ
\id{a}{b} 
= \metric{a}{b} + \news{}{a} \news{*b}{} +\newt{}{a} \newt{*b}{}
\endEQ
such that $\news{}{a} \propto \duvol{ab}{} \tfvec{b}$ 
is hypersurface orthogonal to $\bcsurface$, 
with $\{ \news{*a}{},\newt{*a}{} \}$ denoting 
a basis for $T(\partial\Sigma)^\perp$ that is dual to 
$\{ \news{}{a},\newt{}{a} \}$. 
In particular, 
$\news{*a}{} \news{}{a} =\newt{*a}{}\newt{}{a} =1$,
and $\news{*a}{}\newt{}{a} = \newt{*a}{}\news{}{a} =0$. 
This leads to a corresponding decomposition of the spacetime metric
\EQ\label{gdecomp}
\g{ab}{} = \metric{ab}{} +\news{}{a}\news{*}{a} + \newt{}{a} \newt{*}{b}
\endEQ
with $\news{*}{a} =\g{ab}{}\news{*b}{}$
and  $\newt{*}{a} =\g{ab}{}\newt{*b}{}$. 
Now, define a projection operator $\Pr{\bcsurface}$ 
with respect to $\bcsurface$ by 
\EQ\label{newhsid}
\newh{a}{b} = \id{a}{b} - \news{}{a} \news{*b}{}
\endEQ
satisfying 
\EQ
\newh{a}{b} \news{}{b} = 0, \newh{a}{b} \news{*a}{}=0 . 
\endEQ
Then
\EQ\label{hdecomp}
\newh{ab}{} = \g{ab}{} - \news{}{a}\news{*}{a}
= \metric{ab}{} +\newt{}{a}\newt{*}{a}
\endEQ
defines the induced metric on $\bcsurface$. 
Also, define the volume form on $\bcsurface$ by 
\EQ
\vol{abc}{}(\newh{}{}) = \vol{abcd}{}(g) \news{*d}{}
= 3 \newt{}{[a} \vol{bc]}{} . 
\endEQ
Finally, note that 
\EQs
&&
\duvol{ab}{} = 4\newt{}{[a} \news{}{b]} ,\quad
\vol{abcd}{} = 3\vol{[ab}{}\duvol{cd]}{} =4\vol{[abc}{}(h) \news{}{d]} ,
\label{perpvolid}\\
&&
\perptfvec{a} = -N \newt{*a}{} ,\quad
\perptfvec{a}  \newt{}{a} = -N ,\quad 
\perptfvec{a} \news{}{a} = \perptfvec{a} \vol{ab}{} =0 , 
\endEQs
for some scalar function $N$. 
This yields the identities
\EQs
&& 
\tfvec{a} = -N \newt{*a}{} +\N{a}{} ,
\\
&&
\tfvec{a} \vol{}{bc} \vol{abcd}{}(g)
= \tfvec{a} \duvol{ad}{} 
= -2N \news{}{d} ,
\label{mainid}\\
&&
\tfvec{a} \vol{}{bc} \vol{abc}{}(h) = -2N ,\quad
\Pr{\partial\Sigma}( \tfvec{a} \vol{abc}{}(\newh{}{}) )
= -N\vol{bc}{} .
\label{mainvolid}
\endEQs
These will be important in the analysis of \bdc/s
for both the gravitational field and matter fields. 

Now we introduce an orthonormal frame for $\g{ab}{}$ given by 
\EQ\label{fr}
\frame{\mu}{a} 
= \metric{a}{\mu} +\news{}{a} \news{*\mu}{} +\newt{}{a} \newt{*\mu}{}
\endEQ
where $\metric{a}{\mu} = \metric{a}{b} \frame{\mu}{b}$ 
is an \onfr/ for $\metric{ab}{}$,
with the coefficients
\EQ\label{frdualst}
\news{*\mu}{} = \news{*a}{} \frame{\mu}{a} ,\
\newt{*\mu}{} = \newt{*a}{} \frame{\mu}{a} 
\endEQ
defined to satisfy
$\mp\newt{*\mu}{}\newt{*\nu}{} \pm \news{*\mu}{}\news{*\nu}{} 
= diag(-1,1,0,0)$ 
if $\tfvec{a}$ is timelike or spacelike, 
or $2\news{*(\mu}{}\newt{*\nu)}{} = diag(-1,1,0,0)$
if $\tfvec{a}$ is null. 
Consequently, 
the frame components of 
$\news{a}{},\newt{a}{},\metric{}{ab},\g{}{ab}$ are given by 
\EQs
&&
\news{\mu}{} =\news{a}{} \frame{\mu}{a} ,\
\newt{\mu}{} =\newt{a}{} \frame{\mu}{a} ,\
\label{frst}\\
&&
\metric{}{\mu\nu} = \metric{}{ab} \frame{\mu}{a} \frame{\nu}{b}
= diag(0,0,1,1) ,\
\flat{}{\mu\nu} = \g{}{ab} \frame{\mu}{a} \frame{\nu}{b} 
=diag(\mp 1,\pm 1,1,1) , 
\label{froth}
\endEQs
where $\{ \news{\mu}{},\newt{\mu}{} \}$ are dual to 
$\{ \news{*}{\mu}=\flat{\mu\nu}{} \news{*\nu}{},
\newt{*}{\mu}=\flat{\mu\nu}{} \newt{*\nu}{} \}$. 
Hereafter we fix the frame coefficients \eqrefs{frdualst}{frst}
to be independent of the spacetime metric $\g{ab}{}$, 
so therefore under a variation $\delta\g{ab}{}$, 
the induced variations 
$\delta\news{}{a}$, $\delta\newt{}{a}$, $\delta\metric{a}{\mu}$, 
$\delta\frame{\mu}{a}$ 
satisfy 
\EQs
&& 
\delta\frame{\mu}{a} 
= \delta\metric{a}{\mu} +\news{*\mu}{} \delta\news{}{a} 
+\newt{*\mu}{} \delta\newt{}{a} 
\\&&
\delta\news{}{a} = \news{}{\mu}\delta\frame{\mu}{a} ,\quad
\delta\newt{}{a} = \newt{}{\mu}\delta\frame{\mu}{a} ,\quad
\delta\metric{a}{\mu} = \metric{\nu}{\mu} \delta\metric{a}{\nu} ,
\\&&
\delta\metric{ab}{} = 2\metric{\mu\nu}{}\frame{\mu}{(a}\delta\frame{\nu}{b)} 
= \metric{(a}{\mu}\delta\metric{b)\mu}{} ,\quad
\delta\g{ab}{} = 2\flat{\mu\nu}{}\frame{\mu}{(a}\delta\frame{\nu}{b)} .
\endEQs
Note, by hypersurface orthogonality of $\news{}{a}$, 
it also follows that
\EQ\label{newhvarsid}
\Pr{\bcsurface}( \delta\news{}{a} ) =0 ,\quad
\Pr{\partial\Sigma}( \delta\newt{}{a} ) =0 ,\quad
\Prperp{\partial\Sigma}( \delta\metric{ab}{} ) =0 . 
\endEQ

Let 
\EQ\label{newfrdecomp}
\newh{a}{\mu} =\newh{a}{b} \frame{\mu}{b}
= \frame{\mu}{a} - \news{}{a} \news{*\mu}{}
\endEQ
which yields a decomposition of the frame with respect to $\bcsurface$,
satisfying 
\EQ
\Pr{\partial\Sigma}( \newh{a}{\mu} ) = \metric{a}{\mu} ,\quad
\Prperp{\partial\Sigma}( \newh{a}{\mu} ) = \newt{}{a} \newt{*\mu}{} .
\endEQ
It is convenient for later to also introduce a fixed frame 
adapted to $\partial\Sigma$ and $\tfvec{a}$. 
Let 
\EQ\label{newadaptedfr}
\adfr{0}{a} =\t{}{a} ,\
\adfr{1}{a} =\s{}{a} ,\
\adfr{2}{[a} \adfr{3}{b]} =\vol{ab}{} ,\
\adfr{2}{a} =\vol{a}{b} \adfr{3}{b}  , 
\endEQ
which defines the frame $\adfr{\mu}{a}$
uniquely up to rotations of $\adfr{2}{a},\adfr{3}{a}$. 
Thus, in this formalism, 
$\adfr{\mu}{a}$ is an orthonormal frame 
when $\tfvec{a}$ is timelike or spacelike,
and a null frame when $\tfvec{a}$ is null. 

In the case when $\perptfvec{a}$ is timelike, 
the previous formalism reduces to that in \Ref{Anco-TungI}. 
Most important, the formalism here applies equally well to the cases
when $\perptfvec{a}$ is spacelike or null. 

Finally, in the case that $\perptfvec{a}=0$, 
\ie/ $\tfvec{a}$ is tangential to $\partial\Sigma$, 
we simply fix any basis $\{ \news{}{a},\newt{}{a} \}$ 
of $T^*(\partial\Sigma)^\perp$
and define a frame $\frame{\mu}{a}$ to satisfy the previous equations
\eqsref{fr}{froth}. 
This yields the same formalism as in the case that 
$\tfvec{a}$ is not tangential to $\partial\Sigma$,
except that there does not exist a hypersurface $\bcsurface$ 
generated by $\perptfvec{a}=0$. 

Now, with the frame $\frame{\mu}{a}$ 
used as the gravitational field variable, 
the Lagrangian for the vacuum Einstein equations is given by 
\EQ\label{Lfr}
\L{abcd}(\theta) = \vol{abcd}{}(\theta)\scurv(\theta) . 
\endEQ
Here 
$\scurv(\theta) 
=  \frame{b}{\nu} \curv{b}{\nu}(\theta)$
and 
$\curv{a}{\mu}(\theta) 
= \frame{b}{\nu} \curv{ab}{\mu\nu}(\theta)$
are the scalar curvature and Ricci curvature 
obtained from the curvature 2-form
\EQ\label{frcurv}
\curv{ab}{\mu\nu}(\theta) 
= 2\der{[a}\frcon{b]}{\mu\nu}(\theta)
+ 2\frcon{[a}{\mu\sigma}(\theta)\frcon{b]\sigma}{\nu}(\theta)
\endEQ
in terms of the frame-connection given by 
\EQ\label{frcon}
\frcon{a}{\mu\nu}(\theta) 
= \frame{b\mu}{} \covgder{a}\frame{\nu}{b}
= 2\frame{b[\mu}{} \der{[a} \frame{\nu]}{b]} 
- \frame{b\mu}{}\frame{c\nu}{} \frame{}{a\alpha}\der{[b}\frame{\alpha}{c]} . 
\endEQ
A variation of $\L{abcd}(\theta)$ yields
\EQ
\delta\L{abcd}(\theta)
= \E\mixedindices{\mu}{[bcd}(\theta) \delta\frame{\mu}{a]}
+ \der{[a} \T{bcd]}(\theta,\delta\theta)
\endEQ
where 
\EQ\label{frfieldeq}
\E\mixedindices{\mu}{bcd}(\theta)
= 8 \vol{bcda}{}(\theta) ( 
\curv{}{a\mu}(\theta) -\frac{1}{2}\frame{a\mu}{}\scurv(\theta) )
=0
\endEQ
are the vacuum Einstein field equations for $\frame{\mu}{a}$,
and where
\EQ\label{Tfr}
\T{bcd}(\theta,\delta\theta) 
= 8\vol{abcd}{}(g) \frame{e}{\nu}\frame{a}{\mu} 
\delta\frcon{e}{\mu\nu}(\theta)
\endEQ
is the symplectic potential 3-form.
It follows that 
the Noether current associated to $\tfvec{a}$ is given by the 3-form
\EQ
\J{abc}(\xi,\theta)
= \T{abc}(\theta,\Lie{\xi}\theta) +4\tfvec{d}\L{abcd}(\theta)
= \vol{abcd}{}(g) ( 8\frame{e}{\mu}\frame{d}{\nu} 
\Lie{\xi}\frcon{e}{\mu\nu}(\theta) 
+ 4\tfvec{d} \scurv(\theta) )
\label{Jfr}
\endEQ
which simplifies to 
\EQ
\J{abc}(\xi;\theta)
= 3 \der{[a} \Q{bc]}(\xi;\theta) 
-\tfvec{e} \frame{}{e\mu} \E\mixedindices{\mu}{abc}(\theta)
\endEQ
where
\EQ\label{Qfr}
\Q{bc}(\xi;\theta) 
= 4\tfvec{e} \vol{bcda}{}(g) \frame{d}{\mu} \frame{a}{\nu}
\frcon{e}{\mu\nu}(\theta) 
\endEQ
is the Noether charge potential. 

The gravitational Noether charge associated to $\partial\Sigma$ is
determined by the pullback of $\Q{bc}(\xi;\theta)$. 
A simple expression for the pullback is obtained 
through identities \eqrefs{duvolid}{perpvolid}, 
yielding
\EQ\label{newQfr}
\vol{}{bc} 
\tfvec{a} \vol{debc}{}(g) \frame{d}{\mu} \frame{e}{\nu}
\frcon{a}{\mu\nu}(\theta) 
= \duvol{de}{} \frame{d}{\mu} \frame{e}{\nu}
\tfvec{a} \frcon{a}{\mu\nu}(\theta)
= 4\newt{}{\mu} \news{}{\nu} \tfvec{a} \frame{e\mu}{}
\covgder{a}\frame{\nu}{e}
= 4\tfvec{a} \news{}{\nu} \newt{e}{} \covgder{a}\frame{\nu}{e} 
\endEQ
where, recall, $\vol{bc}{}$ is the volume form on $\partial\Sigma$.
Hence, the surface integral
\EQ\label{totalQg}
Q_\Sigma(\xi;\theta)
= \int_{\partial\Sigma} \Q{bc}(\xi;\theta) 
= 8 \int_{\partial\Sigma} 
\vol{bc}{}  \news{}{\nu} \newt{e}{} \tfvec{a}\covgder{a}\frame{\nu}{e} 
\endEQ
gives the gravitational Noether charge. 

When $\tfvec{a}$ is not tangential to $\partial\Sigma$,
the pullback of $\T{bcd}(\theta,\delta\theta)$ 
to the hypersurface $\bcsurface$ can be simplified similarly by
the identities \eqrefs{mainid}{perpvolid}
and the frame decomposition \eqref{newfrdecomp},
\EQs
\frac{1}{8}\vol{}{ab} \tfvec{c} \T{abc}(\theta,\delta\theta)
&&
= \vol{}{ab} \tfvec{c} \vol{abcd}{}(g) 
\frame{d}{\mu} \frame{e}{\nu} \delta\frcon{e}{\nu\mu}(\theta) 
\nonumber\\&&
= -2N \news{}{\mu} \frame{e}{\nu} 
\delta( \frame{c\nu}{} \covgder{e}\frame{\mu}{c} )
\nonumber\\&&
= -2N \newh{\nu}{e} \delta( \news{}{\mu} \newh{e}{d} \newh{\nu}{c}
\covgder{d}\frame{\mu}{c} )
\nonumber\\&&
= \vol{}{ab} \tfvec{c} \vol{abc}{}(\newh{}{}) 
\newh{\nu}{e} \delta\newK{e}{\nu}
\label{newsimpleTfr}
\endEQs
where we define
\EQ
\newK{a}{\mu} 
= \news{}{\nu} \newh{a}{d} \newh{}{c\mu} \covgder{d}\frame{\nu}{c} . 
\endEQ
(Note these expressions have the same form 
as those obtained in \Ref{Anco-TungI}
when $\tfvec{a}$ is timelike.)
Hence, one obtains
\EQ
\Pr{\bcsurface} \T{abc}(\theta,\delta\theta) 
= 8 \vol{abc}{}(\newh{}{}) \newh{\nu}{e} \delta\newK{e}{\nu} .
\endEQ
This expression leads to a simple form for 
the gravitational symplectic flux
associated to $\bcsurface$,
\EQs
&&
\int_{\bcsurface} \w{abc}(\theta,\delta_1\theta,\delta_2\theta) 
= \int_{\bcsurface} \delta_1\T{abc}(\theta,\delta_2\theta) 
- \delta_2\T{abc}(\theta,\delta_1\theta) 
\nonumber\\&&
= 8 \int_{\bcsurface} \vol{abc}{}(h) \bigg(
( \delta_1\h{\mu}{d} - \h{\mu}{d} \h{e}{\nu} \delta_1\h{\nu}{e} )
\delta_2\K{d}{\mu}
- ( \delta_2\h{\mu}{d} - \h{\mu}{d} \h{e}{\nu} \delta_2\h{\nu}{e} )
\delta_1\K{d}{\mu}
\bigg) . 
\endEQs
The vanishing of this flux determines the allowed boundary conditions
on the frame $\frame{\mu}{c}$ at the boundary hypersurface $\bcsurface$. 

We remark that in a frame \eqref{newadaptedfr} adapted to $\partial\Sigma$,
one sees that 
$\K{a}{\mu} = \h{}{c\mu} \h{a}{d}\covgder{d} \news{}{c}$ 
represents the frame components of the extrinsic curvature tensor 
$\K{ab}{} = \h{a}{d}\h{b}{c}\covgder{d} \news{}{c}$ 
of the boundary hypersurface $\bcsurface$. 
Moreover, the Noether charge \eqref{totalQg} is simply 
$Q_\Sigma(\xi;\theta)
= 8 \int_{\partial\Sigma} 
\vol{bc}{}  \tfvec{a} \newt{}{\mu} \K{a}{\mu} 
= 8 \int_{\partial\Sigma} \vol{bc}{} \tfvec{a} \newt{e}{}\K{ae}{}$.

For the sequel, we now introduce 
Dirichlet and Neumann symplectic vectors 
\EQs
\P{\rm D}{a}(\theta) && 
= \news{}{\nu} \newt{c}{} \metric{a}{d} \covgder{d}\frame{\nu}{c}
-\news{}{\nu} \newt{}{a} \metric{}{bd} \covgder{b}\frame{\nu}{d}
+ \newt{}{\nu} \news{}{a} \metric{}{bd} \covgder{b}\frame{\nu}{d} 
= \frac{1}{2} \vol{}{bc} \vol{acde}{} \frame{d}{\mu} \frame{e}{\nu} 
\frcon{b}{\mu\nu}(\theta) ,
\label{newPDfr}\\
\P{\rm N}{a}(\theta) && 
= \news{}{\nu} \newt{c}{} \covgder{a}\frame{\nu}{c} 
= \frac{1}{4} \vol{}{bc} \vol{bcde}{} \frame{d}{\mu} \frame{e}{\nu}
\frcon{a}{\mu\nu}(\theta) , 
\label{newPNfr}
\endEQs
associated to the boundary 2-surface $\partial\Sigma$ 
and the frame $\frame{\mu}{a}$. 
In a frame \eqref{newadaptedfr} adapted to the hypersurface $\bcsurface$, 
these vectors take the more geometrical form
\EQs
&& 
\P{\rm D}{a}(\vartheta) 
= \newt{c}{} \metric{a}{d} \covgder{d}\news{}{c}
-\newt{}{a} \metric{}{bd} \covgder{b}\news{}{d}
+ \news{}{a} \metric{}{bd} \covgder{b}\newt{}{d} , 
\label{newPD}\\
&& 
\P{\rm N}{a}(\vartheta) 
= \newt{c}{} \covgder{a}\news{}{c} . 
\label{newPN}
\endEQs
Similarly to the derivation in \Ref{Anco-TungI} 
holding for the situation when $\bcsurface$ is timelike, 
here the projection of the vectors \eqrefs{newPDfr}{newPNfr} 
along $\tfvec{a}$ 
yields the respective boundary terms 
required to define a covariant Hamiltonian
for the vacuum Einstein equations 
with $\tfvec{a}$ as the time-flow vector field in a spacetime region
with spatial boundary 2-surface $\partial\Sigma$,
subject to Dirichlet or Neumann boundary conditions 
on the frame $\frame{\mu}{a}$, 
for a timelike, spacelike, or null boundary hypersurface $\bcsurface$.
The significance and properties of the full vectors 
\eqrefs{newPDfr}{newPNfr} will be discussed in \secref{properties}.

Lastly, we make some remarks on the gauge invariance of 
the preceeding results, which follow from the detailed 
gauge transformation analysis given in Sec.~3 of \Ref{Anco-TungI}. 
Under a local $SO(3,1)$ transformation on the frame $\frame{\mu}{a}$,
the Noether charge $\Q{bc}(\xi;\theta)$ transforms inhomogeneously
due to its explicit dependence on the frame connection. 
However, the curvature $\curv{ab}{\mu\nu}(\theta)$ is invariant,
and consequently so is the Lagrangian $\L{abcd}(\theta)$.
Therefore the symplectic current 
$\w{abc}(\theta,\delta_1\theta,\delta_2\theta)$ 
is necessarily gauge invariant. 
As a result, up to addition of a locally constructed exact 2-form,
the symplectic current obtained here for 
the frame formulation of the vacuum Einstein equations
must agree with the analogous current derived from 
the standard metric formulation. 
This means that the presymplectic forms 
$\Omega_\Sigma(\theta,\delta\theta,\Lie{\xi}\theta)$
and $\Omega_\Sigma(g,\delta g,\Lie{\xi} g)$
in the two formulations differ by only a boundary term 
(\ie/ a locally constructed 2-form 
integrated over the 2-surface $\partial\Sigma$). 
Correspondingly, the Dirichlet and Neumann symplectic vectors 
associated to $\Omega_\Sigma(g,\delta g,\Lie{\xi} g)$
in the metric formulation are found to be the same as 
the ones given here for the frame formulation,
up to certain gradient terms. 
Furthermore, if $\tfvec{a}$ is timelike and orthogonal to $\partial\Sigma$,
then these gradient terms can be shown to vanish.
In this situation, 
$\tfvec{a}\P{\rm D}{a}(\vartheta)$ 
and $\tfvec{a}\P{\rm N}{a}(\vartheta)$ are precisely
the Dirichlet and Neumann boundary terms 
in the covariant Hamiltonian 
determined by the metric formulation of the vacuum Einstein equations
(see \Ref{Wald-Iyer2} for a discussion of this Hamiltonian). 
Consequently, as noted in \Ref{Anco-TungI}, we find that the expression 
$\tfvec{a}\P{\rm D}{a}(\vartheta)$ reduces to the boundary term
derived by Brown and York \cite{Brown-York1,Brown-York2}
in the standard canonical formalism,
with Dirichlet boundary conditions on the canonical variables 
in the case of a hypersurface boundary $\bcsurface$
where $\tfvec{a}$ is timelike. 
In comparison, the covariant formulation we have presented here
applies equally well when $\tfvec{a}$ is null or spacelike.

\subsection{ Electromagnetic field }

We start by considering a free electromagnetic field $\A{a}$ 
on $(M,\g{ab}{})$, coupled to the gravitational field, 
generalizing the Minkowski background spacetime 
considered in Sec.~2 in \Ref{Anco-TungI}. 
The Lagrangian 4-form for $\A{a}$ is given by 
\EQ\label{MELg}
\L{abcd}(A;\theta)
=\frac{1}{2} \vol{abcd}{}(g) \F{mn} \coF{mn}
= 3\F{[ab} \duF{cd]}
\endEQ
where $\F{ab}=\covgder{[a} \A{b]}= \der{[a} \A{b]}$ 
is the electromagnetic field strength
and $\duF{ab}=\vol{abcd}{}(g)\coF{cd}$ is the dual field strength
in terms of $\coF{cd}=\g{}{ca} \g{}{db} \F{bd}$, 
with $\g{ab}{}=\frame{\mu}{a} \frame{\nu}{b} \flat{\mu\nu}{}$. 
A useful fact here is that $\covgder{a}$ reduces to $\der{a}$
in any skew derivative expression on $M$. 
By variation of $\A{a}$ and $\frame{\mu}{a}$ in this Lagrangian, 
one obtains 
\EQ\label{MEvarLg}
\delta \L{abcd}(A;\theta)
= \vol{abcd}{}(g) ( \covgder{m}( \delta\A{n} \coF{mn} ) 
-\delta\A{n} \covgder{m}\coF{mn} 
- \EMT{\mu}{e}(A;\theta) \delta\frame{\mu}{e} )
\endEQ
where $\EMT{\mu}{e}(A;\theta)\frame{\mu}{e} = \EMT{d}{e}(A;g)$
is the electromagnetic stress-energy tensor given by 
\EQ\label{MEEMTg}
\EMT{a}{b}(A;g)
= 2\F{ac} \coF{bc} -\frac{1}{2} \id{a}{b} \F{mn} \coF{mn} . 
\endEQ
From the coefficient of the variation $\delta\A{a}$ 
in $\delta \L{abcd}(A;\theta)$, 
the field equation for $\A{a}$ is given by the Maxwell equations
\EQ\label{MEfieldeqg}
*\E\downindex{a}(A;g) 
= \covgcoder{b} \F{ba} = \covgcoder{b} \der{[b} \A{a]} =0 . 
\endEQ

The symplectic potential 3-form obtained from $\L{abcd}(A;\theta)$
is given by 
the total derivative term in \Eqref{MEvarLg}, 
which yields
\EQ\label{METg}
\T{bcd}(A,\delta A;\theta)
= 4\vol{abcd}{}(g) \delta\A{e} \coF{ae} . 
\endEQ
Hence, the Noether current associated to $\tfvec{a}$ for $\A{a}$ 
is given by the 3-form
\EQs
\J{abc}(\xi,A;g) && 
=\T{abc}(A,\Lie{\xi}A;\theta)+ 4\tfvec{d}\L{abcd}(A;\theta)
\nonumber\\&&
= \vol{abcd}{}(g) ( -4\coF{de} \Lie{\xi}\A{e} +2\tfvec{d} \F{mn}\coF{mn} )
\nonumber\\&&
= 4\vol{dabc}{}(g) ( \tfvec{e} \EMT{e}{d}(A;g) 
+\tfvec{e}\A{e} *\E\upindex{d}(A;g) )
+ 6\der{[a}( \duF{bc]} \tfvec{e} \A{e} )
\label{MEJg}
\endEQs
by a similar derivation as in Minkowski spacetime,
with
\EQ 
\Lie{\xi} \A{a} 
= \tfvec{e} \covgder{e} \A{a} +\A{e}\covgder{a} \tfvec{e}
= 2\tfvec{e} \covgder{[e} \A{a]} +\covgder{a}( \tfvec{e} \A{e} ) . 
\endEQ
This yields the electromagnetic Noether charge
\EQ
Q_\Sigma(\xi;A) 
= \int_\Sigma \J{abc}(\xi,A;g)
= 4 \int_\Sigma \tfvec{e} \vol{dabc}{}(g) \EMT{e}{d}(A;g)
+ 2 \int_{\partial\Sigma} \vol{bcda}{} \coF{da} \tfvec{e} \A{e}
\endEQ
for solutions $\A{a}$ of the Maxwell equations \eqref{MEfieldeqg}. 

The total Lagrangian for the Maxwell equations 
coupled to the Einstein equations 
\EQ
\curv{\mu}{a}(\theta) -\frac{1}{2} \frame{\mu}{a} \scurv(\theta)
= \EMT{\mu}{a}(A;\theta)
\endEQ
using the field variables $\A{a}$ and $\frame{\mu}{a}$ 
is given by 
$\L{abcd}(\theta,A)
= \L{abcd}(\theta) - \L{abcd}(A;\theta)$
from \Eqrefs{Lfr}{MELg}. 
One then obtains the total Noether current
\EQs
\J{abc}(\xi,\theta,A)
&&
= \J{abc}(\xi,\theta) - \J{abc}(\xi,A;\theta)
\nonumber\\&&
= 8\vol{dabc}{}(g) \tfvec{e} \frame{\mu}{e}
( \curv{\mu}{d}(\theta) -\frac{1}{2} \frame{d}{\mu} \scurv(\theta)
- \EMT{\mu}{d}(A;\theta) )
+ 3 \der{[a} \Q{bc]}(\xi,\theta,A)
\endEQs
where 
\EQ
\Q{bc}(\xi,\theta,A)
= \tfvec{a} \vol{bcde}{}(g) ( 
4 \frame{d}{\mu} \frame{e}{\nu} \frcon{a}{\mu\nu}(\theta) 
- 2 \coF{de} \A{a} )
\endEQ
is the Noether charge potential. 
Hence, on solutions of the coupled Einstein-Maxwell equations,
the total Noether charge is given by the surface integral
\EQ
Q_\Sigma(\xi;\theta,A)
= \int_{\partial\Sigma} \Q{bc}(\xi,\theta,A) . 
\endEQ
The electromagnetic part of this expression 
simplifies through identities \eqrefs{duvolid}{perpvolid}, 
yielding
\EQ\label{QApart}
\vol{}{bc} \duF{bc} \tfvec{d} \A{d}
= \duvol{bc}{} \coF{bc} \tfvec{d} \A{d}
= 4\newt{}{b} \news{}{c} \coF{bc} \tfvec{d} \A{d} . 
\endEQ
Then, substituting \Eqref{newQfr} for the gravitational part, 
one obtains
\EQ\label{MEQbt}
Q_\Sigma(\xi;\theta,A)
= \int_{\partial\Sigma} \vol{bc}{}  \tfvec{a}
( 8\news{}{\nu} \newt{e}{} \covgder{a}\frame{\nu}{e}
-4 \newt{}{d} \news{}{e} \coF{de} \A{a} ) . 
\endEQ

The Noether current gives a Hamiltonian conjugate to $\tfvec{a}$ 
on $\Sigma$ under compact support variations 
$\delta\frame{\mu}{a}$ and $\delta\A{a}$,
\EQ
H(\xi;\theta,A)
= 8 \int_\Sigma \vol{dabc}{}(g) \tfvec{e} \frame{\mu}{e}
( \curv{\mu}{d}(\theta) - \frac{1}{2} \frame{d}{\mu} \scurv(\theta)
- \EMT{\mu}{d}(A;\theta) )
\endEQ
up to a boundary term \eqref{MEQbt}. 
For variations $\delta\frame{\mu}{a}$ and $\delta\A{a}$
with support on $\partial\Sigma$, 
after taking into account boundary terms, 
one has 
\EQ\label{MEvarH}
\delta H(\xi;\theta,A)
=  \int_{\partial\Sigma} \delta \Q{ab}(\xi,\theta,A)
- \tfvec{c} \T{abc}(\theta,A,\delta\theta,\delta A)
\endEQ
for Einstein-Maxwell solutions, 
where
\EQs
\T{abc}(\theta,A,\delta\theta,\delta A)
&&
= \T{abc}(\theta,\delta\theta) - \T{abc}(A,\delta A;\theta)
\nonumber\\&&
= \vol{dabc}{}(g) ( 
8\frame{d}{\mu} \frame{e}{\nu} \delta\frcon{e}{\mu\nu}(\theta)
- 4 \coF{de} \delta\A{e} )
\endEQs
is the total symplectic potential 3-form
from \Eqrefs{Tfr}{METg}. 
The electromagnetic part of 
the symplectic potential terms in the Hamiltonian variation \eqref{MEvarH}
can be simplified similarly to expression \eqref{newsimpleTfr} 
for the gravitational part, 
yielding
\EQs
\frac{1}{4} \vol{}{ab} \tfvec{c} \T{abc}(A,\delta A;\theta)
&&
= \vol{}{ab} \tfvec{c} \vol{dabc}{}(g) \coF{de} \delta\A{e} 
\nonumber\\&&
= 2N \news{}{d} \coF{de} \delta\A{e} 
\nonumber\\&&
= -\vol{}{ab} \tfvec{c} \vol{abc}{}(h) \news{}{d} \coF{de} \delta\A{e} 
\label{simpleTA}
\endEQs
through identities \eqrefs{mainid}{mainvolid}.
Thus, one obtains
\EQs
\Pr{\partial\Sigma}( \tfvec{c} \T{abc}(\theta,A,\delta\theta,\delta A) )
&& 
= \Pr{\partial\Sigma}( \tfvec{c} \vol{abc}{}(\newh{}{}) (
8 \newh{\nu}{e} \delta\newK{e}{\nu} 
+4 \news{}{d} \coF{de} \delta\A{e} ) ) 
\label{simpleTtotal}\\
&&
= -N\vol{ab}{}( 8 \newh{\nu}{e} \delta\newK{e}{\nu} 
+4 \news{}{d} \coF{de} \delta\A{e} ) . 
\nonumber
\endEQs

Hence, 
for existence of a Hamiltonian conjugate to $\tfvec{a}$ on $\Sigma$,
there must exist a locally constructed 3-form $\B{abc}(\theta,A)$ 
such that
\EQ\label{Beq}
\Pr{\partial\Sigma}( \tfvec{c} \T{abc}(\theta,A,\delta\theta,\delta A) )
= \Pr{\partial\Sigma}( \tfvec{c} \delta\B{abc}(\theta,A)
- \der{[a} \a{}{b]}(\xi,\theta,A,\delta\theta,\delta A) )
\endEQ
for some locally constructed 1-form 
$\a{}{b}(\xi,\theta,A,\delta\theta,\delta A)$ in $T^*(\partial\Sigma)$. 
Then the total Hamiltonian is given by 
$H(\xi;\theta,A)$ plus a boundary term
\EQ\label{newHbtA}
H_B(\xi;\theta,A)
= \int_{\partial\Sigma} \Q{ab}(\xi,\theta,A) - \tfvec{c} \B{abc}(\theta,A) . 
\endEQ

We now consider Dirichlet and Neumann type \bdc/s 
on the fields $\frame{\mu}{a}$ and $\A{a}$ at $\partial\Sigma$. 

First, consider the case when $\tfvec{a}$ is tangential to $\partial\Sigma$. 
Then one finds 
$\Pr{\partial\Sigma}( \tfvec{c}\T{abc}(\theta,A,\delta\theta,\delta A) )
=0$, 
which leads to the following result. 

\Proclaim{ Proposition 2.1. }{
Suppose $\tfvec{a}$ is tangential to $\partial\Sigma$. 
Then {\rm no} \bdc/s are necessary for 
existence of a Hamiltonian conjugate to $\tfvec{a}$ on $\Sigma$. 
Consequently, a Hamiltonian is given by 
$H_\Sigma(\xi;\theta,A)= H(\xi;\theta,A)+ Q_\Sigma(\theta,A)$. }

Next, assume $\tfvec{a}$ is not tangential to $\partial\Sigma$,
and consider Dirichlet and Neumann boundary conditions 
on the electromagnetic and gravitational field variables. 

\Proclaim{ Theorem 2.2. }{
Suppose $\tfvec{a}$ is nowhere tangential to $\partial\Sigma$. 
Let 
\EQs
({\rm D})\qquad && 
\delta( \newh{a}{b} \A{b} )|_{\partial\Sigma} =0 ,\quad
\delta( \newh{a}{\mu} )|_{\partial\Sigma} =0
\\
({\rm N})\qquad &&
\delta( |\newh{}{}| \news{}{b} \newh{c}{a} \coF{cb} )|_{\partial\Sigma} =0 ,
\quad
\delta( \newK{a}{\mu} )|_{\partial\Sigma} =0
\endEQs
where $|\newh{}{}|=\det( \newh{a}{\mu} )$
is the determinant of the components of the frame $\newh{a}{\mu}$ 
associated to $\bcsurface$.
Under Dirichlet (D) or Neumann (N) \bdc/s
for both $\A{a}$ and $\frame{\mu}{a}$,
there exists a Hamiltonian $H(\xi;\theta,A)+H_B(\xi;\theta,A)$
conjugate to $\tfvec{a}$ on $\Sigma$,
with the boundary term \eqref{newHbtA} given by 
\EQs
&&
\H{D}(\xi;\theta,A) 
= 8 \int_{\partial\Sigma} \tfvec{a}( \P{\rm D}{a}(\theta) 
- \frac{1}{2} \newt{}{d} \news{}{e} \coF{de} \A{a} ) dS , 
\label{MEDbt}\\
&&
\H{N}(\xi;\theta,A) 
= 8 \int_{\partial\Sigma} \tfvec{a}( \P{\rm N}{a}(\theta) 
- \metric{[a}{c} \newt{}{d]} \news{}{e} \coF{de} \A{c} ) dS ,
\label{MENbt}
\endEQs 
in terms of the Dirichlet and Neumann symplectic vectors
\eqrefs{newPDfr}{newPNfr}. }

\proclaim{ Proof: }

For case (D), first note from \Eqref{simpleTA} that 
$\vol{}{bc} \tfvec{a} \T{abc}(A,\delta A;\theta)
= 8N \news{}{d} \coF{de} \h{e}{c}\delta\A{c}$.
Now, using the \bdc/ (D) on $\delta\A{a}$,
one has
\EQ
\newh{e}{c} \delta\A{c} 
= \newh{e}{c} \delta( \news{}{c} \news{*b}{}\A{b} )
= \news{*b}{}\A{b} \newh{e}{c} \delta\news{}{c} 
=0
\endEQ
by the hypersurface orthogonality relations \eqrefs{newhsid}{newhvarsid}.
Thus,
\EQ\label{bcDeqA}
\Pr{\partial\Sigma}( \tfvec{a} \T{abc}(A,\delta A;\theta) ) =0 .
\endEQ

Then, in \Eqref{newsimpleTfr}, 
since $\delta\vol{abc}{}(\newh{}{})=\delta\vol{bc}{}=0$
by the \bdc/ (D) on $\delta\frame{\mu}{a}$, 
one has
\EQ
\vol{}{bc} \tfvec{a} \T{abc}(\theta,\delta\theta) 
= \vol{}{bc} \delta( 8\tfvec{a} \vol{abc}{}(\newh{}{}) \newtrK )
\endEQ
and thus,
\EQ\label{bcDeqfr}
\Pr{\partial\Sigma}( \tfvec{a} \T{abc}(\theta,\delta\theta) 
- \delta( 8\tfvec{a} \vol{abc}{}(\newh{}{}) \newtrK ) )
=0
\endEQ
where $\newtrK= \newh{\nu}{e} \newK{e}{\nu}$. 
Hence, substitution of \Eqrefs{bcDeqA}{bcDeqfr} 
into \Eqref{Beq} yields
\EQ
\tfvec{a} \B{abc}(\theta,A)
= 8 \tfvec{a} \vol{abc}{}(\newh{}{}) \newtrK 
\endEQ
and $\a{}{b}=0$. 
This leads to the boundary term \eqref{MEDbt} through \Eqref{newHbtA}
as follows. 
The pullback of $\tfvec{a} \B{abc}(\theta,A)$ to $\partial\Sigma$ 
is given by
\EQ
8 \tfvec{a} \vol{}{bc} \vol{abc}{}(\newh{}{}) \newtrK 
= 8 \tfvec{a} \newt{}{a} 
\news{}{\nu} \newh{d}{e} \covgder{e}\frame{d\nu}{}
= 16 \tfvec{a} \newt{}{a} \news{}{\nu} \newh{d}{e} \covgder{e}\frame{d\nu}{}
\endEQ
which, when combined with expression \eqref{newQfr} 
for the pullback of $\Q{bc}(\xi,\theta)$, yields
\EQs
\vol{}{bc}( \Q{bc}(\xi,\theta) - \tfvec{a} \B{abc}(\theta,A) ) 
&&
= 16 \tfvec{a} ( \news{}{\nu} \newt{e}{} \covgder{a} \frame{\nu}{e}
- \newt{}{a}\news{}{\nu} \newh{d}{e} \covgder{e} \frame{d\nu}{} )
\nonumber\\&&
= 16 \tfvec{a} \news{}{\nu} ( 
\newt{d}{} \metric{a}{e} - \newt{}{a} \metric{}{de} )
\covgder{e} \frame{\nu}{d} 
\nonumber\\&&
= 16 \tfvec{a} \P{\rm D}{a}(\theta)
\endEQs
by the metric decompositions \eqrefs{gdecomp}{hdecomp}
and the orthogonality $\tfvec{a} \news{}{a} =0$. 
Finally, combining this expression with 
the pullback of $\Q{bc}(\xi,A)$ given by \Eqref{QApart},
we obtain \Eqref{MEDbt}.

For case (N), one has from \Eqref{simpleTA} that
\EQs
\vol{}{bc} \tfvec{a} \T{abc}(A,\delta A;\theta) 
&&
= \vol{}{bc} \delta( 4\tfvec{a} \vol{abc}{}(h) \news{}{e} \A{d} \coF{de} )
+4 \vol{}{bc} \A{e} \tfvec{a} \delta( 
\vol{abc}{}(\newh{}{}) \news{}{d}\coF{de} ) . 
\endEQs
Now, since 
\EQ
\delta( \vol{abc}{}(\newh{}{}) \news{}{d} \coF{de} )
= \vol{abc}{}(\newh{}{})
( \delta( \news{}{d} \newh{m}{e} \coF{dm} ) 
+\delta\ln|\newh{}{}| \news{}{d}  \newh{m}{e} \coF{dm} ) , 
\endEQ
this term vanishes by \bdc/ (N) for $\delta\coF{ab}$, 
and thus
\EQ\label{bcNeqA}
\Pr{\partial\Sigma}( \tfvec{a} \T{abc}(A,\delta A;\theta) 
- \delta( 4\tfvec{a} \vol{abc}{}(h) \news{}{e} \A{d} \coF{de} ) )
=0 . 
\endEQ

Next, 
$\Pr{\partial\Sigma}( \tfvec{a} \T{abc}(\theta,\delta\theta) ) =0$
holds immediately by \bdc/ (N) for $\delta\newK{a}{\mu}$. 
Hence, from \Eqrefs{Beq}{bcNeqA}, one has $\a{}{b}=0$ and 
\EQ
\tfvec{a} \B{abc}(\theta,A)
= 4 \tfvec{a} \vol{abc}{}(h) \news{}{e} \A{d} \coF{de} . 
\endEQ
Then this leads to the boundary term \eqref{MENbt} through \Eqref{newHbtA}
similarly to the derivation of the boundary term \eqref{MEDbt} above. 
\endproof

\subsection{ Klein-Gordon scalar field }

We next consider a free \KG/ scalar field $\kg{}$ 
coupled to the gravitational field on $(M,\g{ab}{})$, 
with the standard Lagrangian 4-form given by 
\EQ\label{KGLg}
\L{abcd}(\kg{};\theta)
=\frac{1}{2} \vol{abcd}{}(g) ( 
\covgder{e}\kg{} \covgcoder{e}\kg{} + m^2 \kg{}{}^2 )
\endEQ
where $m=\const$ is the mass. 
Note, here 
$\covgder{a}\kg{}= \der{a}\kg{}$, 
$\covgcoder{a}\kg{} = \g{}{ab} \der{b}\kg{}$,
and $\g{ab}{}=\frame{\mu}{a} \frame{\nu}{b} \flat{\mu\nu}{}$. 
A variation of this Lagrangian with respect to $\kg{}$ and $\frame{\mu}{a}$
yields
\EQ\label{KGvarLg}
\delta \L{abcd}(\kg{};\theta)
= \vol{abcd}{}(g) ( \covgder{e}( \delta\kg{} \covgcoder{e}\kg{} )
-\delta\kg{} ( \covgcoder{e}\covgder{e}\kg{} -m^2 \kg{} )
- \EMT{\mu}{e}(\kg{};\theta) \delta\frame{\mu}{e} )
\endEQ
where $\EMT{\mu}{e}(\kg{};\theta) \frame{\mu}{e}= \EMT{d}{e}(\kg{};g)$
is the \KG/ stress-energy tensor given by 
\EQ\label{KGEMTg}
\EMT{a}{b}(\kg{};g)
= \covgcoder{b}\kg{} \covgder{a}\kg{} 
-\frac{1}{2} \id{a}{b} ( 
\covgder{e}\kg{} \covgcoder{e}\kg{} + m^2 \kg{}{}^2 ) . 
\endEQ
Hence, from the coefficient of the variation $\delta\kg{}$ 
in \Eqref{KGvarLg}, 
the \KG/ field equation for $\kg{}$ is given by 
\EQ\label{KGfieldeqg}
*\E(\kg{};g) 
= \covgcoder{a}\der{a}\kg{} - m^2 \kg{} =0 . 
\endEQ

The symplectic potential 3-form obtained from $\L{abcd}(\kg{};\theta)$
is given by 
\EQ\label{KGTg}
\T{bcd}(\kg{},\delta\kg{};\theta)
= 4\vol{abcd}{}(g) \covgcoder{a}\kg{} \delta\kg{} . 
\endEQ
This yields the Noether current associated to $\tfvec{a}$ 
\EQs
\J{abc}(\xi,\kg{};g) 
&& 
=\T{abc}(\kg{},\Lie{\xi}\kg{};\theta)+ 4\tfvec{d}\L{abcd}(\kg{};\theta)
\nonumber\\&&
= \vol{abcd}{}(g) ( -4\covgcoder{d}\kg{} \Lie{\xi}\kg{} 
+2\tfvec{d}( \covgder{m}\kg{} \covgcoder{m}\kg{} + m^2 \kg{}{}^2 )
\nonumber\\&&
= 4 \vol{dabc}{}(g) \tfvec{e} \EMT{e}{d}(\kg{};g) 
\label{KGJg}
\endEQs
where 
$\Lie{\xi} \kg{} = \tfvec{e} \covgder{e}\kg{}
= \tfvec{e} \der{e}\kg{}$. 
Hence, one obtains the Noether charge
\EQ
Q_\Sigma(\xi;\kg{}) 
= \int_\Sigma \J{abc}(\xi,\kg{};g)
= 4 \int_\Sigma \vol{dabc}{}(g) \tfvec{e} \EMT{e}{d}(\kg{};g) . 
\endEQ
In contrast to the situation for the electromagnetic field,
here, due to the scalar nature of the \KG/ field, 
the Noether charge does not have a surface integral term. 

The total Lagrangian for the \KG/ equation 
coupled to the Einstein equations 
\EQ
\curv{\mu}{a}(\theta) -\frac{1}{2} \frame{\mu}{a} \scurv(\theta)
= \EMT{\mu}{a}(\kg{};\theta)
\endEQ
using the field variables $\kg{}$ and $\frame{\mu}{a}$ 
is obtained through \Eqrefs{Lfr}{KGLg} by 
$\L{abcd}(\theta,\kg{})
= \L{abcd}(\theta) - \L{abcd}(\kg{};\theta)$. 
The resulting total Noether current is given by 
\EQs
\J{abc}(\xi,\theta,\kg{})
&&
= \J{abc}(\xi,\theta) - \J{abc}(\xi,\kg{};\theta)
\nonumber\\&&
= 8\vol{dabc}{}(g) \tfvec{e} \frame{\mu}{e}
( \curv{\mu}{d}(\theta) - \frac{1}{2} \frame{d}{\mu} \scurv(\theta)
- \EMT{\mu}{d}(\kg{};\theta) )
+ 3 \der{[a} \Q{bc]}(\xi,\theta)
\endEQs
where $\Q{bc}(\xi,\theta)$ 
is the gravitational Noether charge potential \eqref{Qfr}. 
Thus, there is no contribution from $\kg{}$ 
to the total Noether charge. 

The Noether current gives a Hamiltonian conjugate to $\tfvec{a}$ 
on $\Sigma$ under compact support variations 
$\delta\frame{\mu}{a}$ and $\delta\kg{}$,
\EQ
H(\xi;\theta,\kg{})
= 8 \int_\Sigma \vol{dabc}{}(g) \tfvec{e} \frame{\mu}{e}
( \curv{\mu}{d}(\theta) - \frac{1}{2} \frame{d}{\mu} \scurv(\theta)
- \EMT{\mu}{d}(\kg{};\theta) )
\endEQ
up to a boundary term 
$\int_{\partial\Sigma} \Q{ab}(\xi,\theta)$. 
For variations $\delta\frame{\mu}{a}$ and $\delta\kg{}$
with support on $\partial\Sigma$, 
one has for Einstein-\KG/ solutions, 
\EQ
\delta H(\xi;\theta,\kg{})
=  \int_{\partial\Sigma} \delta \Q{ab}(\xi,\theta)
- \tfvec{c} \T{abc}(\theta,\delta\theta)
\endEQ
where $\T{abc}(\theta,\delta\theta)$ 
is the expression \eqref{newsimpleTfr}
for the gravitational symplectic potential. 
Thus, there exists a Hamiltonian conjugate to $\tfvec{a}$ on $\Sigma$
if 
\EQ
\Pr{\partial\Sigma}( 
\tfvec{c} \T{abc}(\theta,\kg{},\delta\theta,\delta\kg{}) )
= \Pr{\partial\Sigma}( \tfvec{c} \delta\B{abc}(\theta,\kg{})
- \der{[a} \a{}{b]}(\xi,\theta,\kg{},\delta\theta,\delta\kg{}) )
\endEQ
holds for a locally constructed 3-form $\B{abc}(\theta,\kg{})$ 
and 1-form $\a{}{b}(\xi,\theta,\kg{},\delta\theta,\delta\kg{})$. 
Then the total Hamiltonian is given by 
$H(\xi;\theta,\kg{})$ plus a boundary term
\EQ\label{newHbtkg}
H_B(\xi;\theta,\kg{})
= \int_{\partial\Sigma} \Q{ab}(\xi,\theta) 
- \tfvec{c} \B{abc}(\theta,\kg{}) . 
\endEQ

Now, by an analysis similar to that for the Einstein-Maxwell equations, 
we obtain the following results. 

\Proclaim{ Proposition 2.3. }{
Suppose $\tfvec{a}$ is tangential to $\partial\Sigma$. 
Then {\rm no} \bdc/s are necessary for 
existence of a Hamiltonian conjugate to $\tfvec{a}$ on $\Sigma$. 
Consequently, a Hamiltonian is given by 
$H_\Sigma(\xi;\theta,\kg{})= H(\xi;\theta,\kg{})+ Q_\Sigma(\theta)$. }

\Proclaim{ Theorem 2.4. }{
Suppose $\tfvec{a}$ is nowhere tangential to $\partial\Sigma$. 
Let 
\EQs
({\rm D})\qquad && 
\delta( \kg{} )|_{\partial\Sigma} =0 ,\quad
\delta( \newh{a}{\mu} )|_{\partial\Sigma} =0
\\
({\rm N})\qquad &&
\delta( |\newh{}{}| \news{a}{}\der{a}\kg{} )|_{\partial\Sigma} =0 ,\quad
\delta( \newK{a}{\mu} )|_{\partial\Sigma} =0
\endEQs
where $|\newh{}{}|=\det( \newh{a}{\mu} )$
is the determinant of the components of the frame $\newh{a}{\mu}$ 
associated to $\bcsurface$.
Under Dirichlet (D) or Neumann (N) \bdc/s
for both $\kg{}$ and $\frame{\mu}{a}$,
there exists a Hamiltonian 
$H(\xi;\theta,\kg{})+H_B(\xi;\theta,\kg{})$
conjugate to $\tfvec{a}$ on $\Sigma$,
with the boundary term \eqref{newHbtkg} given by 
\EQs
&&
\H{D}(\xi;\theta,\kg{}) 
= 8 \int_{\partial\Sigma} \tfvec{a} \P{\rm D}{a}(\theta) dS , 
\label{KGDbt}\\
&&
\H{N}(\xi;\theta,\kg{})
= 8 \int_{\partial\Sigma} \tfvec{a}( \P{\rm N}{a}(\theta) 
- \frac{1}{2} \newt{}{a} \kg{} \news{d}{} \der{d}\kg{} ) dS , 
\label{KGNbt}
\endEQs 
in terms of the Dirichlet and Neumann symplectic vectors
\eqrefs{newPDfr}{newPNfr}. }

\subsection{ Yang-Mills and Higgs fields }

Last, we generalize the previous two examples by considering 
on $(M,\g{ab}{})$
a set of \YM/ fields $\ymA{\Upsilon}{a}$ and Higgs fields $\kg{\Upsilon}$, 
$\Upsilon=1,\ldots,n$, 
with a gauge group given by 
any $n$-dimensional semi-simple Lie group $\cal G$, $n\geq 3$. 
Let $\C{\Delta\Lambda}{\Upsilon}$ be the commutator structure constants
of the Lie algebra $\cal A$ of $\cal G$ (in a fixed basis). 
The structure constants are skew
$\C{\Delta\Lambda}{\Upsilon} = -\C{\Lambda\Delta}{\Upsilon}$
and satisfy the Jacobi relation 
$\C{[\Delta\Lambda}{\Upsilon} \C{\Pi]\Upsilon}{\Phi}=0$. 
Let 
$\ymconn{a}{\Upsilon}{\Delta}(A) 
= \C{\Lambda\Delta}{\Upsilon}\ymA{\Lambda}{a}$
be the \YM/ connection,
and define 
$\k{\Delta\Upsilon}{}=
-\frac{1}{2} \C{\Delta\Lambda}{\Pi}\C{\Upsilon\Pi}{\Lambda}$, 
which denotes the positive definite Cartan-Killing metric on $\cal A$.

The \YM/ Lagrangian for $\ymA{\Upsilon}{a}$ is given by the 4-form
\EQ\label{YMLg}
\L{abcd}(\ymA{}{};\theta)
=\frac{1}{2} \vol{abcd}{}(g) \k{\Lambda\Upsilon}{} 
\ymF{\Lambda}{mn} \ymcoF{\Upsilon}{mn}
= 3\k{\Lambda\Upsilon}{} \ymF{\Lambda}{[ab} \ymduF{\Upsilon}{cd]}
\endEQ
where 
\EQ
\ymF{\Lambda}{ab}
=\covgder{[a} \ymA{\Lambda}{b]} 
+\frac{1}{2} \C{\Delta\Upsilon}{\Lambda} \ymA{\Delta}{a}\ymA{\Upsilon}{b}
\endEQ
is the \YM/ field strength, 
$\ymduF{\Lambda}{ab}=\vol{abcd}{}(g)\ymcoF{\Lambda}{cd}$ 
is the dual field strength
in terms of $\ymcoF{\Lambda}{cd}=\g{}{ca} \g{}{db} \ymF{\Lambda}{bd}$, 
with $\g{ab}{}=\frame{\mu}{a} \frame{\nu}{b} \flat{\mu\nu}{}$.
For $\kg{\Upsilon}$, it is convenient to introduce 
the gauge-covariant Higgs field strength
\EQ
\W{\Lambda}{a}
=  \covgder{a}\kg{\Lambda} +e \ymconn{a}{\Lambda}{\Upsilon}(A)\kg{\Upsilon} .
\endEQ
In terms of this field strength 
the Higgs Lagrangian is given by the 4-form
\EQ\label{HLg}
\L{abcd}(\kg{};\theta)
= \vol{abcd}{}(g) 
( \frac{1}{2} \k{\Lambda\Upsilon}{} \W{\Lambda}{m} \coW{m}{\Upsilon} 
+ V(|\kg{}|) )
\endEQ
where $V(|\kg{}|)$ is a Higgs potential
with $|\kg{}|^2 = \k{\Lambda\Upsilon}{} \kg{\Lambda}\kg{\Upsilon}$,
and $e=\const$ is a coupling constant. 
These Lagrangians are gauge invariant under \YM/ gauge symmetries
on the fields $\ymA{\Upsilon}{a}$ and $\kg{\Upsilon}$. 
(In particular, 
if $\ymU{\Upsilon}{\Lambda}$ denotes a homomorphism of $\cal A$
given by a function of the spacetime coordinates $\x{\mu}{}$,
it is straightforward to show that the \YM/ gauge symmetry 
is then given by 
$\ymA{\Upsilon}{a} \rightarrow 
\yminvU{\Upsilon}{\Lambda} \ymA{\Lambda}{a} 
-\duC{\Upsilon\Delta}{\Lambda} \yminvU{\Lambda}{\Pi} \der{a}\ymU{\Pi}{\Delta}$
and 
$\kg{\Upsilon} \rightarrow 
\yminvU{\Upsilon}{\Lambda} \kg{\Lambda}$,
where 
$\duC{\Upsilon\Delta}{\Lambda} 
= \k{}{\Upsilon\Pi}\C{\Lambda\Pi}{\Delta}$
denotes the structure constants of the dual Lie algebra ${\cal A}^*$
and $\k{}{\Upsilon\Pi}$ is the inverse of the Cartan-Killing metric.
Under these transformations, 
the field strengths are gauge covariant,
$\ymF{\Upsilon}{ab} \rightarrow 
\yminvU{\Upsilon}{\Lambda}\ymF{\Lambda}{ab}$
and
$\W{\Upsilon}{a} \rightarrow 
\yminvU{\Upsilon}{\Lambda}\W{\Lambda}{a}$.)

To proceed, we consider the combined Yang-Mills-Higgs Lagrangian,
\EQ\label{YMHLg}
\L{abcd}(\ymA{}{},\kg{};\theta)
= \L{abcd}(\ymA{}{};\theta) + \L{abcd}(\kg{};\theta) .
\endEQ
First, the coefficient of the variation $\delta\ymA{\Upsilon}{a}$ yields
the \YM/ field equations
\EQ
\k{}{\Upsilon\Lambda} *\E\downindex{a\mit\Upsilon}(A;g)
=  \covgcoder{b} \ymF{\Lambda}{ba} 
+\ymcoconn{b}{\Lambda}{\Upsilon}(A) \ymF{\Upsilon}{ba} 
-e\C{\Pi\Upsilon}{\Lambda} \kg{\Pi}\W{\Upsilon}{a} 
=0
\endEQ
where $e\C{\Pi\Upsilon}{\Lambda} \kg{\Pi}\W{\Upsilon}{a}$
has the role of a current source. 
The coefficient of the variation $\delta\kg{\Upsilon}$ 
similarly yields 
the Higgs field equations
\EQ
\k{}{\Upsilon\Lambda} *\E\downindex{\mit\Upsilon}(A;g)
= \covgcoder{a} \W{\Lambda}{a} 
+\ymcoconn{a}{\Lambda}{\Upsilon}(A) \W{\Upsilon}{a} 
- \frac{1}{|\kg{}|} V'(|\kg{}|) \kg{\Lambda} 
=0
\endEQ
Next, by variation of $\frame{\mu}{a}$, 
one obtains the \YM/-Higgs stress-energy tensor
$\EMT{\mu}{e}(A,\kg{};\theta)\frame{\mu}{e}= \EMT{d}{e}(A,\kg{};g)$
where 
\EQ\label{YMEMTg}
\EMT{a}{b}(A,\kg{};g)
=  \k{\Lambda\Upsilon}{} (
2 \ymF{\Lambda}{ac} \ymcoF{\Upsilon}{bc}
+ \W{\Lambda}{a} \coW{\Upsilon}{b} )
-\frac{1}{2} \id{a}{b} 
\k{\Lambda\Upsilon}{} ( \ymF{\Lambda}{mn} \ymcoF{\Upsilon}{mn}
+ \W{\Lambda}{m} \coW{m}{\Upsilon} ) 
- \id{a}{b} V(|\kg{}|) . 
\endEQ

The symplectic potential 3-form arising from 
the Lagrangian \eqref{YMHLg} is given by 
\EQ\label{YMTg}
\T{bcd}(A,\kg{},\delta A,\kg{};\theta)
= 4\vol{abcd}{}(g) \k{\Lambda\Upsilon}{} (
\delta\ymA{\Lambda}{e} \ymcoF{\Upsilon}{ae}
+ \delta\kg{\Lambda} \coW{\Upsilon}{a} ) . 
\endEQ
This yields the Noether current
\EQs
\J{abc}(\xi,A,\kg{};g) && 
=\T{abc}(A,\kg{},\Lie{\xi}A,\Lie{\xi}\kg{};\theta)
+ 4\tfvec{d}\L{abcd}(A,\kg{};\theta)
\nonumber\\&&
= \vol{abcd}{}(g) ( 
\k{\Lambda\Upsilon}{} ( 
-4\ymcoF{\Lambda}{de} \Lie{\xi}\ymA{\Upsilon}{e} 
+\tfvec{d} 2\ymF{\Lambda}{mn} \ymcoF{\Upsilon}{mn} )
\nonumber\\&&\qquad
+ \k{\Lambda\Upsilon}{} ( 
-4\coW{\Lambda}{d} \Lie{\xi}\kg{\Upsilon}
+ 2\W{\Lambda}{m} \coW{m}{\Upsilon} ) )
+4 V(|\kg{}|) )
\nonumber\\&&
= 4 \vol{dabc}{}(g) ( \tfvec{e} \EMT{e}{d}(A,\kg{};g) 
+ 6\der{[a}( \k{\Lambda\Upsilon}{} \ymduF{\Lambda}{bc]} 
\tfvec{e} \ymA{\Upsilon}{e} )
\label{YMJg}
\endEQs
for Yang-Mills-Higgs solutions. 
Hence, one obtains the Noether charge
\EQ
Q_\Sigma(\xi;A,\kg{}) 
= \int_\Sigma \J{abc}(\xi,A,\kg{};g)
= 4 \int_\Sigma \tfvec{e} \vol{dabc}{}(g) \EMT{e}{d}(A,\kg{};g)
+ 2\int_{\partial\Sigma} \k{\Lambda\Upsilon}{} \vol{bcda}{}
\ymcoF{\Lambda}{da} \tfvec{e} \ymA{\Upsilon}{e} . 
\endEQ

For the Yang-Mills-Higgs equations
coupled to the Einstein equations 
\EQ
\curv{\mu}{a}(\theta) -\frac{1}{2} \frame{\mu}{a} \scurv(\theta)
= \EMT{\mu}{a}(A,\kg{};\theta)
\endEQ
using the field variables 
$\ymA{\Upsilon}{a}$, $\kg{\Upsilon}$, $\frame{\mu}{a}$, 
the total Lagrangian is given by 
$\L{abcd}(\theta,A,\kg{})
= \L{abcd}(\theta) -\L{abcd}(A,\kg{};\theta)$
from \Eqrefs{Lfr}{YMHLg}. 

Through same analysis as used in the Maxwell and \KG/ examples, 
we obtain the following results. 

\Proclaim{ Proposition 2.5. }{
Suppose $\tfvec{a}$ is tangential to $\partial\Sigma$. 
Then {\rm no} \bdc/s are necessary for 
existence of a Hamiltonian conjugate to $\tfvec{a}$ on $\Sigma$. 
Consequently, a Hamiltonian is given by 
\EQ
H(\xi;\theta,A,\kg{})
= 8 \int_\Sigma \vol{dabc}{}(g) \tfvec{e} \frame{\mu}{e}
( \curv{\mu}{d}(\theta) - \frac{1}{2} \frame{d}{\mu} \scurv(\theta)
- \EMT{\mu}{d}(A,\kg{};\theta) )
\endEQ
up to an inessential boundary term. }

\Proclaim{ Theorem 2.6. }{
Suppose $\tfvec{a}$ is nowhere tangential to $\partial\Sigma$. 
Let 
\EQs
({\rm D})\qquad && 
\delta( \ymA{\Upsilon}{a} )|_{\partial\Sigma} =0 ,\quad
\delta( \kg{\Upsilon} )|_{\partial\Sigma} =0 ,\quad
\delta( \newh{a}{\mu} )|_{\partial\Sigma} =0
\\
({\rm N})\qquad &&
\delta( |\newh{}{}| \news{}{b} \newh{c}{a} \ymcoF{\Upsilon}{cb} )
|_{\partial\Sigma} =0 ,\quad
\delta( |\newh{}{}| \news{a}{}\W{\Upsilon}{a} )|_{\partial\Sigma} =0 ,\quad
\delta( \newK{a}{\mu} )|_{\partial\Sigma} =0
\endEQs
where $|\newh{}{}|=\det( \newh{a}{\mu} )$
is the determinant of the components of the frame $\newh{a}{\mu}$ 
associated to $\bcsurface$.
Under Dirichlet (D) or Neumann (N) \bdc/s
for both $\kg{}$ and $\frame{\mu}{a}$,
there exists a Hamiltonian 
$H(\xi;\theta,A,\kg{}) + H_B(\xi,\theta,A,\kg{})$
conjugate to $\tfvec{a}$ on $\Sigma$,
with the boundary term given by 
\EQs
&&
\H{D}(\xi;\theta,A,\kg{}) 
= 8 \int_{\partial\Sigma} \tfvec{a} 
( \P{\rm D}{a}(\theta) - \P{\rm D}{a}(A) ) dS , 
\label{YMDbt}\\
&&
\H{N}(\xi;\theta,A,\kg{}) 
= 8 \int_{\partial\Sigma} \tfvec{a}
( \P{\rm N}{a}(\theta) - \P{\rm N}{a}(A,\kg{}) ) dS , 
\label{YMNbt}
\endEQs 
where
\EQs
&&
\P{\rm D}{a}(A)
= \frac{1}{2} \k{\Lambda\Upsilon}{} 
\newt{}{d} \news{}{e} \ymcoF{\Upsilon}{de} \ymA{\Lambda}{a} , 
\\&&
\P{\rm N}{a}(A,\kg{})
= \k{\Lambda\Upsilon}{} (
\metric{[a}{c} \newt{}{d]} \news{}{e} \ymcoF{\Upsilon}{de} \ymA{\Lambda}{c} 
+ \frac{1}{2} \newt{}{a} \news{d}{} \W{\Upsilon}{d} \kg{\Lambda} ) , 
\endEQs 
and $\P{\rm D}{a}(\theta)$, $\P{\rm N}{a}(\theta)$
are the symplectic vectors given by \eqrefs{newPDfr}{newPNfr}. }

\subsection{ Remarks }
\label{remarks}

Clearly, the previous results 
when $\tfvec{a}$ is not tangential to $\partial\Sigma$
are easily generalized to 
mixed Dirichlet-Neumann \bdc/s on the tetrad and matter fields
similar to Theorem~2.4 and Theorems~3.5 and~3.6 in \Ref{Anco-TungI}. 
In particular, for allowed \bdc/s, 
note that one can have 
the tetrad satisfying (D) while the matter fields satisfy (N),
and vice versa.

\section{ Properties of the symplectic vectors }
\label{properties}

We first review some geometry of spatial 2-surfaces in spacetime
(most of this material is standard, \eg/, \cite{ONeill,Epp,Szabados}). 
Then we describe the properties of 
the Dirichlet and Neumann symplectic vectors
regarded as locally constructed geometrical vector fields
associated to a fixed spatial 2-surface in spacetime,
independently of any Hamiltonian structure.

\subsection{2-surface geometry}

Let $(S,\metric{ab}{})$ be 
a closed, orientable smooth spacelike 2-surface 
in a spacetime $(M,\g{ab}{})$, 
where $\metric{ab}{}$ is the pullback of $\g{ab}{}$ to $S$.
Let $\TS$ and $\TperpS$ denote, respectively, 
the tangent space of $S$ and the normal space to $S$ 
(defined by the orthogonal complement of $\TS$ in $\TM$). 
Since $\TS\oplus \TperpS=\TM$, 
every vector in $\TM$ has a unique decomposition into vectors
tangent and normal to $\TS$,
given by projection operators 
$\PrS :\TM \rightarrow \TS$, 
$\PrperpS :\TM \rightarrow \TperpS$. 

Fix an oriented \onfr/ $\{\t{a}{},\s{a}{}\}$ for $\TperpS$,
\EQ
\t{a}{} \s{}{a} =0, -\t{a}{}\t{}{a}=\s{a}{}\s{}{a}=1, 
\endEQ
with $\t{a}{}$ being a future timelike unit vector
and $\s{a}{}$ being an outward spacelike unit vector. 
(If $M$ is spatially non-compact, 
we define the ``outward'' direction by the exterior of the set $M-S$. 
If $M$ is spatially compact, 
there is no preferred way in general to distinguish the sets $S$ and $M-S$,
so we then make an arbitrary consistent choice for an ``outward'' direction. )
The metric on $S$ is given by 
\EQ
\metric{ab}{} =\g{ab}{} +\t{}{a}\t{}{b} -\s{}{a}\s{}{b} . 
\endEQ
The compatible volume form on $S$ is given by 
\EQ
\vol{ab}{}=\vol{abcd}{}(g) \s{c}{}\t{d}{} ,
\endEQ
satisfying
$\metric{c[a}{}\metric{b]d}{} =\vol{ab}{}\vol{cd}{}$. 
The projection operators for $\TS$ and $\TperpS$ are given by 
\EQ
\PrSop{a}{b}
= \metric{a}{b}=\vol{ac}{}\vol{}{bc}, 
\PrperpSop{a}{b}= \s{}{a}\s{b}{}-\t{}{a}\t{b}{} =\perpmetric{a}{b}
= \perpvol{ac}{}\perpvol{}{bc}
\endEQ
where $\perpvol{ab}{}=2\s{}{[a}\t{}{b]}$
and 
\EQ
\perpmetric{ab}{} =\s{}{a}\s{}{b} -\t{}{a}\t{}{b} . 
\endEQ

Both $\metric{ab}{}$ and $\vol{ab}{}$ are independent of 
choice of the \onfr/. 
Since $\perpmetric{ab}{}$ is a two-dimensional Lorentz metric, 
any two oriented \onfr/s 
$\{\t{a}{},\s{a}{}\}$ and $\{\bt{a}{},\bs{a}{}\}$ 
differ by a local boost
\EQ\label{frboost}
\bt{a}{}=(\cosh\ha) \t{a}{} + (\sinh\ha) \s{a}{}, 
\bs{a}{}=(\cosh\ha) \s{a}{} + (\sinh\ha) \t{a}{},
\endEQ
where $\ha$ is a function on $S$. 
Under an arbitrary boost \eqref{frboost},
$\metric{ab}{}$ and $\vol{ab}{}$ are invariant. 

The intrinsic geometry of the 2-surface $S$ is completely determined by 
the metric $\metric{ab}{}$. 
In particular, the intrinsic curvature of $S$ is given by 
\EQ
[\D{a},\D{b}] \v{}{c} = \curvS{abc}{d}\v{}{d}
\endEQ
where $\v{}{c}$ is any dual tangent vector field on $S$,
and $\D{a}$ denotes the metric compatible (torsion-free) 
derivative operator on $S$ defined by $\D{a}\metric{bc}{}=0$. 
Since $S$ is two-dimensional, it follows that 
the intrinsic curvature tensor has only one linearly independent component
\EQ
\curvS{abc}{d} = \frac{1}{2}\metric{c[a}{}\metric{b]}{d} \scurvS
= \frac{1}{2}\vol{ab}{}\vol{c}{d} \scurvS
\endEQ
where $\scurvS$ denotes the scalar curvature of $S$. 

The 2-surface $S$ also has an extrinsic geometry 
with respect to $(M,\g{ab}{})$,
which is characterized by the following curvatures 
\cite{ONeill,Chen}.
Let $\covSder{a} =\metric{a}{b}\covder{b}$
where $\covder{b}$ is 
the metric compatible (torsion-free) derivative operator on $(M,\g{ab}{})$. 
Then $\covSder{a}$ can be decomposed into 
the tangential derivative operator $\D{a}$
and a normal derivative operator $\perpD{a}$, 
with $\covSder{a}=\D{a} +\perpD{a}$,
defined by 
$\perpD{a}\v{b}{} = \perpmetric{c}{b}\covSder{a}\v{c}{}$
for any vector field $\v{a}{}$ in $T(M)$ at $S$. 
Now consider $\covSder{a} \t{}{b}$ and $\covSder{a} \s{}{b}$. 
The tangential parts yield the extrinsic curvature tensors of $S$
with respect to the \onfr/
\EQ\label{excurvs}
\excurv{}{ab}(t) = \D{a} \t{}{b} ,\
\excurv{}{ab}(s) = \D{a} \s{}{b} , 
\endEQ
which are symmetric tensors on $S$. 
These measure the spatial rotation of the \onfr/ in $\TperpS$ 
under displacement on $S$. 
The normal parts of $\covSder{a} \t{}{b}$ and $\covSder{a} \s{}{b}$ 
give 
\EQ
\perpD{a}\t{}{b} = \s{}{b}\norScon{}{a} ,\
\perpD{a}\s{}{b} = -\t{}{b}\norScon{}{a} , 
\endEQ
where 
\EQ
\norScon{}{a}= \s{c}{}\covSder{a} \t{}{c} , 
\endEQ
which measures the boost of the \onfr/ in $\TperpS$ 
under displacement on $S$. 
The commutator of $\perpD{a}$ defines 
the {\it normal curvature} of $S$
\EQ
[\perpD{a},\perpD{b}] \v{}{c} = \norScurv{abc}{d}\v{}{d} , 
\endEQ
with 
\EQ
\norScurv{abcd}{} = 2\D{[a} \norScon{}{b]} \perpvol{cd}{} , 
\endEQ
where $\v{}{c}$ is any dual normal vector field on $S$. 
Hence, 
$\norScon{}{a}$ is geometrically a connection 1-form on $S$
associated to the normal curvature of $S$. 
Since $S$ is two-dimensional, note $\norScurv{abcd}{}$
has only one linearly independent component, 
which is proportional to $\vol{}{ab} \covSder{a} \norScon{}{b}$.

The trace of the extrinsic curvatures \eqref{excurvs} of $S$
\EQ\label{trexcurvs}
\excurv{}{}(t) = \metric{}{ab}\excurv{}{ab}(t) ,\
\excurv{}{}(s) = \metric{}{ab}\excurv{}{ab}(s) , 
\endEQ
measure how the 2-surface area changes 
under infinitesimal dragging of $S$ along each direction of the \onfr/.
In particular, for any vector field $\v{a}{}$ in $\TperpS$,
\EQ\label{expansion}
\PrS( \Lie{v} \vol{ab}{} ) = \excurv{}{}(v)\vol{ab}{} , 
\endEQ
with 
\EQ
\excurv{}{}(v) = \D{a} \v{a}{} 
= \frac{1}{2} \metric{}{ab} \Lie{v} \metric{ab}{} . 
\endEQ
Thus, $S$ is ``expanding'' or ``contracting'' in the direction $\v{a}{}$
according to whether its trace extrinsic curvature $\excurv{}{}(v)$ 
is positive or negative. 
(More precisely, 
$\excurv{}{}(v)$ equals the rate of change of the area of the image of $S$
under any diffeomorphism of $M$ whose generator agrees with $\v{a}{}$ at $S$.)
We say that the expansion of $S$ defined by \eqref{expansion}
for a direction $\v{a}{}$ is spacelike, timelike, or null, 
if $\v{a}{}\v{}{a}$ is, respectively, positive, negative, or zero. 
If $\v{a}{}$ is non-null, 
we refer to 
$\frac{1}{|v|} |\excurv{}{}(v)|$
as the {\it absolute expansion} of $S$ in the direction $\v{a}{}$
(with $|v|=\sqrt{|\v{a}{}\v{}{a}|}$).

A preferred direction in $\TperpS$ is given by the mean curvature vector
\cite{ONeill,Chen} 
\EQ\label{meancurvvec}
\mcurvH{a}{} = \excurv{}{}(s) \s{a}{} -  \excurv{}{}(t) \t{a}{} . 
\endEQ
If $\mcurvH{a}{}$ is spacelike or timelike, 
then this is the direction of, respectively, 
minimum absolute spacelike or minimum absolute timelike 
expansion of $S$. 
Furthermore, the minimum value of the absolute expansion 
is given by the mean extrinsic curvature of $S$, 
$\frac{1}{|\mcurvH{}{}|} |\trexcurv(\mcurvH{}{})|
= \sqrt{ |\excurv{}{}(s)^2 - \excurv{}{}(t)^2| }$. 
Note, here, the norm of $\mcurvH{a}{}$ is 
$\mcurvH{a}{}\mcurvH{}{a} = \excurv{}{}(s)^2 - \excurv{}{}(t)^2 
\equiv \mcurvH{2}{}$. 

The mean curvature vector and normal curvature tensor of $S$
are each independent of choice of the \onfr/,
namely, $\mcurvH{a}{}$ and $\norScurv{abcd}{}$ 
are invariant under boosts \eqref{frboost} of  $\{\t{a}{},\s{a}{}\}$.
In contrast, 
the extrinsic curvatures of $S$ are not invariant
but instead transform like the \onfr/,
while the normal connection transforms like a SO$(1,1)$ connection
\EQ
\norScon{'}{a}= \norScon{}{a}+\covSder{a}\ha
\endEQ
with respect to the SO$(1,1)$ group generated by the boosts \eqref{frboost}.

\subsection{Dirichlet Symplectic vector}

It is convenient to work with a null frame for $\TperpS$.
Let
\EQ
\sqrt{2}\outnframe{}{a}= \t{}{a} +\s{}{a}, 
\sqrt{2}\innframe{}{a}= \t{}{a} -\s{}{a}, 
\endEQ
which respectively define 
outgoing and ingoing future pointing null dual vectors
satisfying
\EQ
\outnframe{}{a} \innframe{a}{} = -1 . 
\endEQ
Note that any two such oriented null frames 
$\{ \outnframe{}{a}, \innframe{}{a} \}$
and $\{ \nframe{'+}{a}, \nframe{'-}{a} \}$
are related by a local boost
\EQ\label{nfrboost}
\nframe{'+}{a} = e^{\ha} \nframe{+}{a} ,\
\nframe{'-}{a} = e^{-\ha} \nframe{-}{a} ,
\endEQ
where $\ha$ is a boost parameter given by a function on $S$.

The extrinsic curvatures of $S$ in the $\nframe{\pm}{a}$ directions are
given by 
\EQ
\excurv{\pm}{ab} = \D{a} \nframe{\pm}{b} . 
\endEQ
Then 
\EQ\label{nexcurvs}
\excurv{\pm}{} = \D{a} \nframe{\pm a}{} 
= \frac{1}{2} \metric{}{ab}\Lie{\nframe{\pm}{}} \metric{ab}{}
\endEQ
are the trace extrinsic curvatures
which measure the expansion of $S$ in the $\nframe{\pm}{a}$ directions. 
Specifically, 
$\excurv{\pm}{}$ is the rate of change of 2-surface area of $S$ 
\EQ
\PrS( \Lie{\nframe{\pm}{}} \vol{ab}{} )= \excurv{\pm}{}\vol{ab}{}
\endEQ
under any diffeomorphism of $M$ whose generator is given by 
$\nframe{\pm}{a}$ at $S$.
Equivalently, 
$\excurv{\pm}{}$ measures the focusing of 
a congruence of null geodesics normal to $S$. 

The mean curvature vector of $S$ is given by 
\EQ\label{meancurv}
\mcurvH{a}{} = -(\excurv{-}{} \outnframe{a}{}+\excurv{+}{} \innframe{a}{})
\endEQ
and the connection for the normal curvature of $S$ is given by 
\EQ
\norScon{}{a} =  \outnframe{b}{}\covSder{a} \innframe{}{b}
\endEQ

We now consider the Dirichlet symplectic vector \eqref{newPD} 
associated to $S$ in the frame $\{ \t{a}{},\s{a}{} \}$, 
and separate it into vectors that are normal and tangential to $S$
\EQs
&& \Pperp{}{a}{} = \PrperpS( \P{a}{} )
= \excurv{+}{} \innframe{a}{} - \excurv{-}{} \outnframe{a}{} , 
\label{DnormalP}\\
&& \Ppar{}{a}{} = \PrS( \P{a}{} )
= \metric{}{ac} \innframe{b}{} \covSder{c} \outnframe{}{b} . 
\label{DtangentialP}
\endEQs

We call $\Pperp{}{a}{}$ the {\it Dirichlet normal vector} associated to $S$.
It has the important property that it is independent of choice of 
the null frame at $S$
\cite{Anco-Wong,Epp,Szabados}.

\Proclaim{ Proposition 3.1. }{
$\Pperp{}{a}{}$ is invariant under arbitrary boosts \eqref{nfrboost}
of the oriented null frame. 
}
Consequently, 
$\Pperp{}{a}{}$ depends only on 
the 2-surface $S$ and the spacetime metric $\g{ab}{}$.
(In particular, its components in any coordinate system can be 
locally constructed out of the components of $g$ and their partial derivatives,
but not in a coordinate invariant form.)
Moreover, $\Pperp{}{a}{}$ has three significant geometrical properties.

First of all, 
we consider the extrinsic curvature of $S$ 
in the direction $\Pperp{}{a}{}$
\EQ
\excurv{\perp}{ab} = \D{a} \Pperp{}{}{b} . 
\endEQ
Remarkably, the trace of this extrinsic curvature vanishes
\EQs
\excurv{\perp}{} = \D{a} \Pperp{}{a}{} 
= \excurv{+}{} \D{a}\innframe{a}{} - \excurv{-}{} \D{a}\outnframe{a}{}
=0 
\endEQs
by \Eqref{nexcurvs} and $\nframe{\pm a}{}\covSder{a}=0$. 
Then we have 
\EQ
\Lie{P_\perp} \vol{ab}{} 
= 
\excurv{+}{}\Lie{\innframe{}{}} \vol{ab}{}
-\excurv{-}{}\Lie{\outnframe{}{}} \vol{ab}{}
= \excurv{\perp}{}\vol{ab}{}
=0 . 
\endEQ
This result yields the following key geometrical property of $\Pperp{}{a}{}$.

\Proclaim{ Theorem 3.2. }{
The normal direction $\Pperp{}{a}{}$ to $S$ in the spacetime $(M,\g{ab}{})$
is area preserving, \ie/
$S$ has zero expansion in the direction $\Pperp{}{a}{}$.
}
Moreover, this property essentially characterizes 
the directional part of $\Pperp{}{a}{}$
since there is a unique area-preserving normal direction 
at all points of $S$, except, if any, where 
$\excurv{+}{}=\excurv{-}{}=0$
(in which case all normal directions to $S$ are area-preserving). 

Second, we find that the norm of $\Pperp{}{a}{}$ is given by
\EQ
\Pperp{}{2}{} = 2\excurv{+}{}\excurv{-}{} . 
\endEQ
Hence, the direction of $\Pperp{}{a}{}$ is timelike, spacelike, or null
if the expansions $\excurv{\pm}{}$ of $S$ are, respectively,
opposite sign, same sign, or at least one is zero.
(These are boost invariant properties.)
In general, 
the signs of $\excurv{\pm}{}$ can vary on $S$
even if the spacetime curvature satisfies positive energy conditions. 
Therefore, the sign of $\Pperp{}{2}{}$ need not be the same everywhere on $S$. 
We note that the situation $\Pperp{}{2}{}>0$ characterizes $S$ as 
a trapped surface, related to the formation of black-holes
\cite{Hayward}. 
Further remarks on this aspect of $\Pperp{}{a}{}$ 
will be made in \secref{conclusion}.

Third, we see that $\Pperp{}{a}{}$ is closely related to 
the mean curvature vector of $S$. 

\Proclaim{ Proposition 3.3. }{
\EQ\label{HPorthogonality}
\Pperp{}{a}{} \mcurvH{}{a} =0 ,\
\Pperp{}{2}{}=- \mcurvH{2}{} .
\endEQ
}
Thus, $\Pperp{}{a}{}$ is respectively timelike, spacelike, or null
as $\mcurvH{a}{}$ is spacelike, timelike, or null. 
Let
$\normH = \sqrt{|\mcurvH{2}{}|}$, 
$\normPperp = \sqrt{|\Pperp{}{2}{}|}$
denote the absolute norms of $\mcurvH{a}{}$ and $\Pperp{}{a}{}$.
Then, in the non-null case,
the relations \eqref{HPorthogonality}
give a unique characterization of $\Pperp{}{a}{}$ (up to a sign)
as a vector in $\TperpS$ 
orthogonal to $\mcurvH{a}{}$ 
and with the same absolute norm as $\mcurvH{a}{}$.
Consequently, we will write $\Pperp{}{a}{} = \perpmcurvH{a}{}$
and refer to 
\EQ\label{perpmeancurv}
\perpmcurvH{a}{} = \excurv{+}{} \innframe{a}{} - \excurv{-}{} \outnframe{a}{}
\endEQ
as the {\it normal mean curvature vector} of $S$, 
with $\perpmcurvH{a}{} \mcurvH{}{a} =0$, 
$\perpmcurvH{2}{}=- \mcurvH{2}{}$. 

\Proclaim{ Lemma 3.4. }{
Suppose $\normH\neq 0$ or equivalently  $\normPperp\neq 0$,
\ie/ $\mcurvH{a}{}$ and $\perpmcurvH{a}{}=\Pperp{}{a}{}$ 
are non-null. 
Then $\{ \frac{1}{\normH}\mcurvH{a}{}, \frac{1}{\normH}\perpmcurvH{a}{} \}$
is an orthonormal frame for $\TperpS$. 
Correspondingly, the pair of vectors 
\EQs
&&
\fixinframe{a}{} \equiv
\frac{1}{\sqrt{2}\normH} (\mcurvH{a}{}+\perpmcurvH{a}{}) 
= \frac{ -\excurv{-}{} }{ \sqrt{|\excurv{+}{}\excurv{-}{}|} } \outnframe{a}{},
\\&&
\fixoutframe{a}{} \equiv
\frac{1}{\sqrt{2}\normH} (\mcurvH{a}{}-\perpmcurvH{a}{}) 
= \frac{ -\excurv{+}{} }{ \sqrt{|\excurv{+}{}\excurv{-}{}|} } \innframe{a}{},
\endEQs
is a null frame for $\TperpS$.
}
Thus, in the non-null case, 
$\mcurvH{a}{}$ and $\perpmcurvH{a}{}=\Pperp{}{a}{}$ 
determine a preferred orthonormal frame 
(or null frame) of $\TperpS$ 
which depends just on the 2-surface $S$ and spacetime metric $\g{ab}{}$. 
Then we can summarize the geometrical properties of these vectors
in terms of the following orthonormal vectors in $\TperpS$, 
\EQs
&& \hatH{a}{} 
= \frac{1}{ \sqrt{2|\excurv{+}{}\excurv{-}{}|} }
( \excurv{-}{} \outnframe{a}{} + \excurv{+}{} \innframe{a}{} ) , 
\\
&& \hatperpH{a}{} 
= \frac{1}{ \sqrt{2|\excurv{+}{}\excurv{-}{}|} }
( \excurv{-}{} \outnframe{a}{} - \excurv{+}{} \innframe{a}{} ) . 
\endEQs
\Proclaim{ Theorem 3.5. }{
Suppose $\excurv{+}{}\excurv{-}{}\neq 0$ on $S$,
\ie/ $\mcurvH{a}{}$ and $\perpmcurvH{a}{}$ are non-null. 
Then:
\sp{0}
(1) The expansion of $S$ is zero in the unique normal direction 
$\hatperpH{a}{}$,
which is spacelike or timelike 
as $\excurv{+}{}\excurv{-}{}$ is positive or negative on $S$. 
\sp{0}
(2) The absolute expansion of $S$ in the orthogonal normal direction 
$\hatH{a}{}$ is $\sqrt{2|\excurv{+}{}\excurv{-}{}|}$.
This is the minimum absolute spacelike expansion 
or minimum absolute timelike expansion 
where $\excurv{+}{}\excurv{-}{}$ is, respectively, 
positive or negative on $S$. 
}

We now turn to consider the geometrical properties of $\Ppar{}{a}{}$. 
To begin, 
$\Ppar{}{a}{}$ can be identified \cite{Epp,Szabados}
with the connection $\norScon{}{a}$
for the normal curvature of $S$, 
in the null frame $\{ \outnframe{}{a}, \innframe{}{a} \}$.  

\Proclaim{ Proposition 3.6. }{
\EQ
\Ppar{}{a}{} = -\metric{}{ab} \norScon{}{b}
\endEQ
where 
\EQ
\norScon{}{a} = \outnframe{c}{} \covSder{a} \innframe{}{c} . 
\endEQ
}
Hence, in contrast to the invariance of $\Pperp{}{a}{}$
under boosts \eqref{nfrboost} of the null frame, 
$\Ppar{}{a}{}$ is not invariant but instead transforms
as a SO$(1,1)$ connection
\EQ\label{DboostPpar}
\Ppar{}{'}{a}= \Ppar{}{}{a}-\covSder{a}\ha
\endEQ
where $\ha$ is a boost parameter given by a function on $S$.
This describes a gauge transformation of $\Ppar{}{a}{}$
associated to the boosts \eqref{nfrboost}
acting on $\TperpS$
as an SO$(1,1)$ gauge group. 
Consequently,
the curl of $\Ppar{}{a}{}$ has the role of the gauge invariant curvature. 

\Proclaim{ Proposition 3.7. }{
\EQ
-\D{[a} \Ppar{}{}{b]} = \frac{1}{4}\norScurv{abcd}{} \perpvol{}{cd}
\endEQ
is invariant under arbitrary boosts \eqref{nfrboost} of the null frame,
where $\norScurv{abcd}{}$ is the normal curvature of $S$. 
}
Thus, 
the curvature $\D{[a} \Ppar{}{}{b]}$ depends only on the 2-surface $S$ 
and the spacetime metric $\g{ab}{}$. 

Interestingly, 
in the case when $\mcurvH{a}{}$ is non-null,
we can use the preferred \onfr/ or null frame given by Lemma~3.4
to gauge-fix $\Ppar{}{a}{}$. 
We introduce 
\EQs
\fixPpar{}{a} &&
= \frac{1}{\mcurvH{2}{}} \perpmcurvH{b}{}\D{a} \mcurvH{}{b}
=  \frac{1}{\perpmcurvH{2}{}} \mcurvH{b}{}\D{a} \perpmcurvH{}{b}
\label{DfixPpar}\\&&
= \Ppar{}{}{a} +\frac{1}{2}\covSder{a}\ln(\excurv{+}{}/\excurv{-}{})
\nonumber
\endEQs
related to $\Ppar{}{}{a}$ by a gauge transformation \eqref{DboostPpar}
with boost parameter $\ha=\frac{1}{2}\ln(\excurv{-}{}/\excurv{+}{})$ on $S$. 

\Proclaim{ Proposition 3.8. }{ 
$\fixPpar{}{a}{}$ is invariant under arbitrary boosts \eqref{nfrboost}
of the oriented null frame. 
}
Consequently, we call $\fixPpar{}{a}$
the {\it invariant Dirichlet tangent vector} associated to $S$. 
In particular, 
$\fixPpar{}{a}$ depends only on the 2-surface $S$ 
and the spacetime metric $\g{ab}{}$.
We now state the main geometrical property of $\fixPpar{}{a}$,
which follows from \Eqref{DfixPpar}. 

\Proclaim{ Theorem 3.9. }{
Suppose $\excurv{+}{}\excurv{-}{}\neq 0$ on $S$,
\ie/ $\mcurvH{a}{}$ and $\perpmcurvH{a}{}$ are non-null. 
Then the boost (with respect to $\TperpS$) 
of the area-preserving unit normal vector $\hatperpH{a}{}$ to $S$ 
under displacement on $S$ 
is a maximum in the direction $\fixPpar{}{a}$. 
By orthogonality of $\mcurvH{a}{}$ and $\perpmcurvH{a}{}$,
this is equivalent to the tangent direction on $S$ in which 
the boost of the unit mean curvature vector $\hatH{a}{}$ 
under displacement on $S$ is a maximum. 
}

Finally, from Propositions~3.1 and~3.8, 
when the mean curvature vector is non-null 
we can define an invariant locally constructed Dirichlet 4-vector 
associated to $S$ by 
\EQ
\fixP{a}{} = \Pperp{}{a}{} +\fixPpar{a}{}
= \excurv{+}{} \innframe{a}{} -  \excurv{-}{} \outnframe{a}{}
+ \metric{}{ac} \innframe{b}{} \covSder{c}\outnframe{}{b}
+ \frac{1}{2} \covSder{a} \ln( \excurv{+}{}/\excurv{-}{} ) . 
\endEQ
Note that this vector depends only on $S$ and $\g{ab}{}$
and is independent of the choice of null frame 
$\{ \outnframe{}{a}, \innframe{}{a} \}$.  
Indeed, in terms of purely geometrical structure associated to $S$,
\EQ
\fixP{a}{} 
= \perpmcurvH{a}{} 
+ \frac{1}{\mcurvH{2}{}} \perpmcurvH{b}{} \coD{a}\mcurvH{}{b}
\endEQ
where $\mcurvH{a}{}$ is the mean curvature vector 
\eqref{meancurv}
and $\perpmcurvH{a}{}$ is the normal mean curvature vector 
\eqref{perpmeancurv}
of $S$.

\subsection{ Neumann symplectic vector }

Finally, we consider the Neumann symplectic vector \eqref{newPN}
associated to $S$, 
\EQ
\P{a}{} = \g{}{ac} \innframe{b}{} \covder{c} \outnframe{}{b} . 
\label{NP}
\endEQ
Notice, first of all, 
the tangential part of $\P{a}{}$ with respect to $S$
\EQ
\Ppar{}{a}{} = \PrS( \P{a}{} )
= \metric{}{ac} \innframe{b}{} \covSder{c} \outnframe{}{b}
\label{NtangentialP}
\endEQ
is identically equal to the tangential part of 
the Dirichlet symplectic vector \eqref{DtangentialP}. 
Hence, similarly to Theorem~3.9,
$\Ppar{}{a}{}$ gives the direction in which 
the boost of the null frame $\nframe{\pm}{a}$
under displacement on $S$ is a maximum. 
In contrast, 
the normal part of $\P{a}{}$ with respect to $S$
\EQ
\Pperp{}{a}{} = \PrperpS( \P{a}{} )
= \perpmetric{}{ac} \innframe{b}{} \covder{c} \outnframe{}{b}
= - \perpmetric{}{ac} \outnframe{b}{} \covder{c} \innframe{}{b}
\label{NnormalP}
\endEQ
involves derivatives of the null frame $\nframe{\pm}{a}$
in normal directions to $S$. 
In particular, 
note through substitution of 
\EQ
\perpmetric{a}{b} = 
-( \outnframe{}{a}\innframe{b}{} + \innframe{}{a}\outnframe{b}{} )
\endEQ
that 
\EQ\label{NPperpfrcomm}
\Pperp{}{a}{} = 
-\outnframe{a}{}\innframe{b}{}\innframe{c}{}
\covder{c}\outnframe{}{b}
+ \innframe{a}{}\outnframe{b}{}\outnframe{c}{}
\covder{c}\innframe{}{b}
= \perpmetric{c}{a} [\innframe{}{},\outnframe{}{}]\upindex{c}
\endEQ
since 
$\outnframe{b}{}\covder{c}\outnframe{}{b}
=\innframe{b}{}\covder{c}\innframe{}{b}=0$. 

\Proclaim{ Proposition 3.10. }{
$\Pperp{}{a}{}$ is the normal part of the commutator 
$[\innframe{}{},\outnframe{}{}]\upindex{a}$ 
of the null frame,
$\Pperp{}{a}{} = \PrperpS [\innframe{}{},\outnframe{}{}]\upindex{a}$.
}
Consequently, 
unlike $\Ppar{}{a}{}$ which is well-defined just given 
the 2-surface $S$ and a null frame $\nframe{\pm a}{}$ of $\TperpS$, 
it is necessary to consider ``nearby'' 2-surfaces $S'$, 
diffeomorphic to $S$, 
to extend the null frame $\nframe{\pm}{a}$ of $\TperpS$ off $S$
so that $\Pperp{}{a}{}$ is well-defined. 

Let $\ncongruenceS$ denote a two-parameter $(\lambda_+,\lambda_-)$
local null congruence of 2-surfaces $S'$ 
diffeomorphic to $S$ in $(M,\g{ab}{})$
with $S_{(0,0)}=S$. 
(The congruence is defined to be ingoing as a function of $\lambda_-$
and outgoing as a function of $\lambda_+$.)
Extend the null frame $\{  \outnframe{a}{}, \innframe{a}{} \}$
off $\TperpS$ to $T(S')^\perp$. 
This extension is unique up to boosts \eqref{nfrboost}. 
Then 
$\Pperp{}{a}{} 
= \perpmetric{}{ac} \innframe{b}{} \covder{c} \outnframe{}{b}$
is a well-defined normal vector at each point on $S$. 
We call $\Pperp{}{a}{}$ the {\it Neumann normal vector} associated to $S$
in a null congruence 
$\ncongruenceS \simeq S\times(\lambda_+,\lambda_-)$. 
It depends, of course, on the congruence but also on the choice of
null frame for $T(\ncongruenceS)^\perp$. 

\Proclaim{ Proposition 3.11. }{
Under boosts \eqref{nfrboost} of the null frame
on the 2-surfaces $\ncongruenceS\simeq S\times(\lambda_+,\lambda_-)$, 
$\Pperp{}{a}{}$ transforms as 
\EQ\label{NboostPperp}
\Pperp{}{'}{a}= \Pperp{}{}{a}-\perpmetric{a}{b}\covder{b}\ha
\endEQ
where $\ha$ is a boost parameter given by a function of 
$(\lambda_+,\lambda_-)$. 
}
By Proposition~3.6, 
$\Ppar{}{a}{}$ has a similar boost transformation property,
which has the geometrical meaning of a SO$(1,1)$ connection
for the normal curvature of $S$. 
This suggests that, geometrically, 
$\P{a}{} = \Pperp{}{a}{}+\Ppar{}{a}{}$ is also related to 
a SO$(1,1)$ connection associated to an extrinsic curvature of $S$. 

Consider the derivative operator $\perpcovder{a}$ defined by 
$\perpcovder{a}\v{b}{} = \perpmetric{c}{b} \covder{a} \v{c}{}$
for any normal vector field $\v{a}{}$ on the 2-surfaces $\ncongruenceS$. 
The commutator of $\perpcovder{a}$ gives the curvature
\EQ
[\perpcovder{a},\perpcovder{b}]\v{}{c}
= \perpcurv{abc}{d} \v{}{d} . 
\endEQ
(Note that, on functions, 
$[\perpcovder{a},\perpcovder{b}] f= 2\covder{[a}\covder{b]}f =0$.)
Clearly, 
$\PrperpS( [\perpcovder{a},\perpcovder{b}] )\v{}{c}
= [\perpD{a},\perpD{b}]\v{}{c} = \norScurv{abc}{d} \v{}{d}$
yields the normal curvature of $S$. 
Hence, $\perpcurv{abcd}{}$ is a generalization of $\norScurv{abcd}{}$,
which we call the {\it sectional curvature normal to} $S$
in the null congruence $\ncongruenceS$. 

\Proclaim{ Proposition 3.12. }{
\EQ
\perpcurv{abcd}{}=2\perpcovder{[a} \perpcon{}{b]} \perpvol{cd}{}
\endEQ
where $\perpcon{}{a} = \outnframe{b}{} \covder{a} \innframe{}{b}$. 
}
Here $\perpcon{}{a}$ is geometrically a connection 1-form 
for $\perpcurv{abcd}{}$. 
In particular, boosts of the null frame act as an SO$(1,1)$ gauge group
on $T(\ncongruenceS)^\perp$ under which $\perpcon{}{a}$ 
transforms as
$\perpcon{'}{a} = \perpcon{}{a} +\covder{a}\ha$
where $\ha$ is a function on 
$\ncongruenceS \simeq S\times(\lambda_+,\lambda_-)$.
Note that the curvature $\perpcurv{abcd}{}$ is invariant under these boosts. 
This leads to the main geometrical result concerning $\P{a}{}$. 

\Proclaim{ Theorem 3.13. }{
In any null congruence of 2-surfaces $\ncongruenceS$, 
$\P{}{a} = 2\perpcon{}{a}$
is a connection 1-form for the sectional curvature normal to $S$,
\EQ
-\perpcovder{[a} \P{}{b]} =\frac{1}{4}\perpcurv{abcd}{}\perpvol{}{cd} . 
\endEQ
}
Thus the curl $\perpcovder{[a} \P{}{b]}$ is invariant under
arbitrary boosts of the null frame on $\ncongruenceS$.
It depends, still, on the choice of null congruence 
$\ncongruenceS \simeq S\times(\lambda_+,\lambda_-)$.

In general, there is no unique null congruence determined just by 
$S$ and $\g{ab}{}$. 
However, a natural choice is given by ingoing and outgoing 
null geodesics congruences $S_{\lambda_\pm}$ through $S$,
with $\ncongruenceS$ defined as 
$(S_{\lambda_+})_{\lambda_-}$ or $(S_{\lambda_-})_{\lambda_+}$ 
corresponding to constructing successive one-parameter 
ingoing and outgoing congruences
\cite{Wald-book}.

If $\ncongruenceS \simeq S\times(\lambda_+,\lambda_-)$
is chosen to be a null geodesic congruence,
then the geodesic equation implies that $\nframe{\pm a}{}$ satisfies
$\nframe{'\pm b}{} \covder{b} \nframe{'\pm a}{} =0$
where $\nframe{'\pm a}{}$ is given by a boost \eqref{nfrboost}
for some function $\ha$ of $(\lambda_+,\lambda_-)$.
Thus, 
\EQ\label{geodesiceq}
\outnframe{b}{} \covder{b} \outnframe{a}{} 
= -\outnframe{a}{} \outnframe{b}{} \covder{b} \ha ,\
\innframe{b}{} \covder{b} \innframe{a}{} 
= \innframe{a}{} \innframe{b}{} \covder{b} \ha . 
\endEQ
This leads to a simplification of $\Pperp{}{a}{}$ 
from \Eqref{NPperpfrcomm}, 
\EQs
\Pperp{}{a}{} 
&&
= \outnframe{a}{}\outnframe{b}{}\innframe{c}{}
\covder{c}\innframe{}{b}
- \innframe{a}{}\innframe{b}{}\outnframe{c}{}
\covder{c}\outnframe{}{b}
\nonumber\\
&&
= - \outnframe{a}{} \innframe{c}{}\covder{c}\ha 
-\innframe{a}{} \outnframe{c}{}\covder{c}\ha . 
\endEQs

\Proclaim{ Proposition 3.14. }{
In a null geodesic congruence, 
$\Pperp{}{a}{} = \perpmetric{}{ab} \covder{b}\ha$, 
and consequently, 
$\PrperpS( \perpcurv{abcd}{} )=0$. 
}
The converse of this result also holds, 
since if the normal part of $\perpcurv{abcd}{}$ vanishes,
then $\Pperp{}{}{a}$ is a gradient 
and hence $\nframe{\pm a}{}$ satisfies 
the geodesic equation \eqref{geodesiceq}
so that the congruence $\ncongruenceS$
arises from null geodesics through $S$. 

Geometrically, the boost function $\ha$ 
in the geodesic equation \eqref{geodesiceq}
is related to the choice of parameterization of 
the null congruence. 
Indeed, we can fix the parameterization in a natural way by 
$\ha=0$, 
which implies a corresponding gauge-fixing of $\Pperp{}{a}{}$,
\EQ
\fixPperp{a}{} =0 . 
\endEQ

To conclude, we remark that the use of ingoing and outgoing null congruences
in defining $\Pperp{}{a}{}$ can be replaced by using 
timelike and spacelike congruences, 
denoted $S_{\lambda_s}$ and $S_{\lambda_t}$, 
through $S$. 
Moreover, if $\g{ab}{}$ has isometries
then it may be possible to fix a unique local two-parameter congruence
$S_{(\lambda_s,\lambda_t)}$ constructed in a natural way from 
the Killing vectors and invariant surfaces of the isometries. 
In general, then $\Pperp{}{a}{}$ is no longer just a gradient. 
This will be illustrated in the examples in the next section. 

Finally, it is important to note that there is no ambiguity 
in $\Pperp{}{a}{}$ appearing in 
the Neumann Hamiltonian \bdt/ \eqref{newPNfr}, 
since this involves only the component of $\Pperp{}{a}{}$ 
in the direction of the time-flow vector, 
which is well-defined using the unique timelike congruence 
through $S$ generated by the time-flow vector on $M$.

\section{Examples}
\label{examples}

We now consider examples for the Dirichlet and Neumann symplectic vectors
described in \secref{properties}.
In particular,
we calculate these vectors and their properties for
spacelike, topologically spherical 2-surfaces in
(A) Minkowski Spacetime,
(B) Spherically Symmetric Spacetimes,
(C) Axisymmetric Spacetimes,
(D) Homogeneous Isotropic Spacetimes,
(E) Asymptotically Flat Spacetimes.

\subsection{Minkowski Spacetime}

Consider a closed orientable spacelike 2-surface $S$
embedded in a spacelike hyperplane in Minkowski spacetime
$(\Rnum^4,\flat{}{})$,
using spherical coordinates
\EQ
\label{ex1}
\flat{ab}{}=
-(dt)_a(dt)_b + (dr)_a(dr)_b + r^2( (d\theta)_a (d\theta)_b
+ \sin^2\theta (d\phi)_a (d\phi)_b ) ,
\endEQ
where $S$ is given by $t =t_0$, $r =R(\theta,\phi)$
for some function $R(\theta,\phi)$ and constant $t_0$.
Fix an orthonormal frame $\adfr{\mu}{a}$ 
adapted to $S$ in $(\Rnum^4,\flat{}{})$ by 
\EQs 
\adfr{0}{a} = (dt)_a ,\quad&&
\adfr{1}{a} = \frac{1}{\psi}\left( (dr)_a -\der{a}R \right) , 
\label{ex1frperpS}\\
\adfr{2}{a} = 
\frac{1}{\mu}\left( r(d\theta)_a +\frac{\der{\theta}R}{r} (dr)_a \right) 
,\quad&&
\adfr{3}{a} =
\frac{\mu}{\psi}\left( r\sin\theta (d\phi)_a 
+\frac{\der{\phi}R}{\mu^2 r\sin\theta} 
( (dr)_a -\der{\theta}R (d\theta)_a ) \right) 
\label{ex1frS}
\endEQs
where
\EQs
&&
\mu=\sqrt{ 1+r^{-2} (\der{\theta}R)^2 } ,\quad
\psi = 
\sqrt{ 1+r^{-2}( (\der{\theta}R)^2 + (\der{\phi}R/\sin\theta)^2 ) } . 
\endEQs
Note the metric associated to $S$ is given in spherical coordinates by
\EQs
\metric{ab}{} = 
&& 
( 1-\psi^{-2} ) (dr)_a (dr)_b
+2\psi^{-2} (dr)\downindex{(a} \der{b)}R 
- \psi^{-2}\der{a}R\ \der{b}R 
\nonumber\\&&
+ r^2( (d\theta)_a (d\theta)_b
+ \sin^2\theta (d\phi)_a (d\phi)_b ) . 
\endEQs
The pullback of $\metric{ab}{}$ to $S$ yields the induced metric on $S$
\EQ
\metric{ab}{}|_S = 
( R^2 +(\der{\theta}R)^2 ) (d\theta)_a (d\theta)_b
+ ( R^2\sin^2\theta +(\der{\phi}R)^2 ) (d\phi)_a (d\phi)_b
+2\der{\theta}R \der{\phi}R (d\theta)_{(a} (d\phi)_{b)} . 
\endEQ
Correspondingly, 
let $\psi_S=\psi|_S 
= \sqrt{ 1+R^{-2}( (\der{\theta}R)^2 + (\der{\phi}R/\sin\theta)^2 ) }$. 

The trace of the extrinsic curvatures of $(S,\metric{ab}{})$
with respect to
the frame on $\TperpS$
\EQ\label{ex1TperpSfr}
\t{}{a} =\adfr{0}{a}|_S 
=(dt)_a ,\
\s{}{a}=\adfr{1}{a}|_S 
= \frac{1}{\psi_S}\left( (dr)_a -\der{a}R \right) 
\endEQ
are respectively 
\EQ
\trexcurv(t) = \metric{}{ab} \covder{a} \t{}{b}
=-( \frcon{2}{02}(\vartheta)+\frcon{3}{03}(\vartheta) )|_S
= 0 
\endEQ
and
\EQ
\trexcurv(s) = \metric{}{ab} \covder{a}
\s{}{b} =-( \frcon{2}{12}(\vartheta) +\frcon{3}{13}(\vartheta) )|_S
= 2( \adfr{a}{1} \adfr{b}{2} \der{[a}\adfr{2}{b]} 
+ \adfr{a}{1} \adfr{b}{3} \der{[a}\adfr{3}{b]} )|_{r=R(\theta,\phi)} , 
\endEQ
calculated in terms of the Ricci rotation coefficients 
\EQ
\frcon{\lambda}{\mu\nu}(\vartheta)
= \adfr{a}{\lambda} \adfr{b\nu}{} \covder{a} \adfr{\mu}{b}
= 2\adfr{a}{\lambda} \adfr{b[\mu}{} \der{[a} \adfr{\nu]}{b]}
- \adfr{b\mu}{} \adfr{c\nu}{} \der{[b} \adfr{}{c]\lambda} . 
\endEQ
Here $\trexcurv(s)$ is the standard Euclidean extrinsic curvature 
of $S$ in $\Rnum^3$ \cite{ONeill}. 
(The explicit expression for $\trexcurv(s)$ as a function of 
the spherical coordinates is lengthy and will be omitted.)

A preferred direction in $T(S)^\perp$ is given by 
the mean curvature vector 
\EQ 
\mcurvH{a}{} = \s{a}{} \trexcurv(s) - \t{a}{} \trexcurv(t) 
= \frac{\trexcurv(s)}{\psi_S} \left( 
(\der{r})^a -\frac{\der{\theta}R}{R^2} (\der{\theta})^a
-\frac{\der{\phi}R}{R^2\sin^2\theta} (\der{\phi})^a \right) , 
\endEQ
which is spacelike.
This vector gives
the direction of the minimum absolute spacelike expansion of $S$ 
in $(\Rnum^4,\flat{}{})$. 
Furthermore, the value of the expansion is 
the mean extrinsic curvature of $S$ given by
\EQ
\frac{1}{\sqrt{H^2}} \trexcurv(H)
= |\trexcurv(s)| = \sqrt{H^2} . 
\endEQ
Note $S$ is convex or concave according to where
the sign of $\trexcurv(s)$ is negative or positive. 

The normal part of the Dirichlet symplectic vector is given by the
normal mean curvature vector
\EQ
\Pperp{\rm D}{a}{} = \perpmcurvH{a}{}
= \t{a}{} \trexcurv(s) - \s{a}{} \trexcurv(t) 
= \trexcurv(s) (\der{t})^a . 
\endEQ
Note that
$\Pperp{\rm D}{a}{}$ is timelike, orthogonal to $\mcurvH{a}{}$,
with the same absolute norm as $\mcurvH{a}{}$.
Most significant,
$\Pperp{\rm D}{a}{}$ gives the direction of zero expansion of $S$.

A preferred orthonormal frame for $T(S)^\perp$ is
\EQ\label{ex1HPfr}
\fixt{a}{}
=\frac{1}{\sqrt{H^2}} \perpmcurvH{a}{} 
=(\der{t})^a |_S ,\quad
\fixs{a}{}
=\frac{1}{\sqrt{H^2}} \mcurvH{a}{} 
= \frac{1}{\psi} (
(\der{r})^a -\frac{\der{\theta}R}{r^2}\ (\der{\theta})^a 
-\frac{\der{\phi}R}{r^2\sin^2\theta}\ (\der{\phi})^a )|_S , 
\endEQ
which depend only on $S$ and $\flat{ab}{}$
but not on the Minkowski frame $\adfr{\mu}{a}$.
In the preferred frame \eqref{ex1HPfr}, 
the tangential part of the Dirichlet symplectic vector is
\EQ
\Ppar{\rm D}{a}{}=\metric{}{ac} \fixt{b}{} \covder{c} \fixs{}{b}
=-\frac{1}{\psi_S} \metric{}{ac} \left(
(\der{r})^b -\frac{\der{\theta}R}{R^2}\ (\der{\theta})^b 
-\frac{\der{\phi}R}{R^2\sin^2\theta}\ (\der{\phi})^b \right)
\covder{c} (dt)_b
= 0 , 
\endEQ
and thus the normal curvature of $S$ is zero.

Hence the complete Dirichlet symplectic vector is
\EQ
\P{a}{} = \Pperp{\rm D}{a}{} + \Ppar{\rm D}{a}{}
= \trexcurv(s) (\der{t})^a , 
\endEQ
which depends only on $S$ and $\flat{ab}{}$.
In particular, it is independent of choice of
the original orthonormal frame \eqrefs{ex1frperpS}{ex1frS} on Minkowski space
and of the normals \eqref{ex1TperpSfr} in $\TperpS$.

To define the Neumann symplectic vector,
it is natural to extend the preferred \onfr/ \eqref{ex1HPfr}
off $S$ by using the obvious isometries of $\flat{ab}{}$.
With respect to this extension,
the normal part of the Neumann symplectic vector
is given by the commutator
\EQ
\Pperp{\rm N}{a}{} =
\PrperpS [\fixt{}{},\fixs{}{}]\upindex{a}
= \frac{1}{\psi_S} \PrperpS \left( 
[(\der{t}),(\der{r})]\upindex{a} 
-\frac{1}{R^2\sin^2\theta} 
[(\der{t}), \sin^2\theta\der{\theta}R\ (\der{\theta}) 
+ \der{\phi}R\ (\der{\phi})]\upindex{a} \right)
= 0 . 
\endEQ
(Alternatively, the same result for $\Pperp{\rm N}{a}{}$
is obtained by extending \eqref{ex1HPfr} off $S$
to the congruence of 2-surfaces
$t=\const$, $r-R(\theta,\phi)=\const$,
which lie in parallel hyperplanes to $S$ and are isometric to $S$.)
Then, since $\Ppar{\rm N}{a}{}=\Ppar{\rm D}{a}{}=0$,
the complete Neumann symplectic vector
in the congruence of 2-surfaces associated to $S$
under isometries of $\flat{ab}{}$
is given by
\EQ
\P{a}{} = \Pperp{\rm N}{a}{} + \Ppar{\rm N}{a}{} =0 . 
\endEQ
Thus the sectional curvature normal to $S$ vanishes,
reflecting the fact that $S$ lies in a flat hyperplane. 

\subsubsection*{ Light cone 2-sphere }

Next, consider a closed orientable spacelike 2-surface $S$ 
embedded in a light cone in $(\Rnum^4,\flat{}{})$. 
Let $u=(t-r)/\sqrt{2}$, $v=(t+r)/\sqrt{2}$, $\theta$, $\phi$
be light cone coordinates 
(\ie/ $r,\theta,\phi$ are spherical coordinates with respect to 
the origin point for the cone), 
with 
\EQ\label{ex1a} 
\flat{ab}{}=
- 2(du)_{(a}(dv)_{b)} + \frac{1}{2} (v-u)^2( 
(d\theta)_a (d\theta)_b  + \sin^2\theta (d\phi)_a (d\phi)_b ) .
\endEQ
Then $S$ is given by $u =u_0$, $v =V(\theta,\phi)$ 
for some function $V(\theta,\phi)$ and constant $u_0$. 
Note that $(du)_a$ and $(dv)_a -\der{a}V$ are, respectively, 
a null normal and spacelike normal for $S$, 
while $(\der{\theta})^a +\der{\theta}V (\der{v})^a$
and $(\der{\phi})^a +\der{\phi}V (\der{v})^a$
are orthogonal tangent vectors on $S$. 

Fix a null frame $\adfr{\mu}{a}$ adapted to $S$ in $(\Rnum^4,\flat{}{})$ by
\EQs
&&
\adfr{0}{a} = (du)_a ,\quad 
\adfr{1}{a} =  \frac{1}{2} \psi^2 (du)_a + (dv)_a -\der{a}V ,
\label{ex1afrperpS}\\&&
\adfr{2}{a} = 
r (d\theta)_a -\frac{\der{\theta}V}{r} (du)_a ,\quad
\adfr{3}{a} = 
r\sin\theta (d\phi)_a -\frac{\der{\phi}V}{r\sin\theta} (du)_a 
\label{ex1afrS}
\endEQs
satisfying 
$\adfr{\mu}{a}\adfr{\nu}{b} \flat{}{ab} =
-2\id{0}{(\mu}\id{1}{\nu)} + \id{2}{\mu}\id{2}{\nu} + \id{3}{\mu}\id{3}{\nu}$,
where 
\EQ
\psi=|dV| 
= r^{-1} \sqrt{ (\der{\theta}V)^2 + (\der{\phi}V/\sin\theta)^2 } .
\endEQ
Note the metric associated to $S$ is given by
\EQ
\metric{ab}{} =
\psi^2 (du)_a (du)_b  -2 (du)_{(a} \left( \der{\theta}V (d\theta)_{b)} 
+\der{\phi}V (d\phi)_{b)} \right) 
+ r^2( (d\theta)_a (d\theta)_b + \sin^2\theta (d\phi)_a (d\phi)_b ) , 
\endEQ
where
\EQ
\metric{ab}{}|_S = \frac{1}{2} (V-u_0)^2 
( (d\theta)_a (d\theta)_b + \sin^2\theta (d\phi)_a (d\phi)_b ) 
\endEQ
yields the induced metric on $S$. 

The trace of the extrinsic curvatures of $(S,\metric{ab}{})$ with
respect to the null frame on $\TperpS$
\EQs
&&
\u{}{a} =\adfr{0}{a}|_S 
= (du)_a , 
\label{ex1aTperpSfr+}\\
&&
\v{}{a} =\adfr{1}{a}|_S 
=  \psi_S^2 (du)_a +(dv)_a
-\der{\theta}V (d\theta)_a-\der{\phi}V (d\phi)_a 
\label{ex1aTperpSfr-}
\endEQs
are respectively
\EQ 
\trexcurv(u) = \metric{}{ab} \covder{a}  \u{}{b} 
=-( \frcon{2}{02}(\vartheta)+\frcon{3}{03}(\vartheta) )|_S
= -\frac{2}{R}
\endEQ
and
\EQs
\trexcurv(v) = \metric{}{ab} \covder{a} \v{}{b} 
&&
=-(  \frcon{2}{12}(\vartheta) +\frcon{3}{13}(\vartheta) )|_S
\nonumber\\&&
= \frac{2}{R} (1+\psi_S^2) 
-\frac{2\der{\theta}R}{R^2} \frac{\cos\theta}{\sin\theta}
-\frac{2\nder{2}{\theta}R}{R^2} 
-\frac{2\nder{2}{\phi}R}{R^2 \sin^2\theta}
\endEQs
where 
\EQ
\psi_S = \frac{1}{\sqrt{2}} \psi|_S
= R^{-1} \sqrt{ (\der{\theta}R)^2 + (\der{\phi}R/\sin\theta)^2 } ,\quad
R= \sqrt{2} r|_S = V-u_0 .
\endEQ

A preferred direction in $T(S)^\perp$ is given by the mean curvature vector
\EQ
\mcurvH{a}{} =-\u{a}{} \trexcurv(v) - \v{a}{} \trexcurv(u)
\endEQ
in terms of the null vectors 
\EQ\label{ex1anullnormal}
\u{a}{} =-(\der{v})^a |_S ,\quad
\v{a}{} = -\left( (\der{u})^a +\frac{\psi^2}{2} (\der{v})^a 
+ \frac{\der{\theta}V}{r^2} (\der{\theta})^a
+\frac{\der{\phi}V}{r^2 \sin^2\theta} (\der{\phi})^a \right)|_S . 
\endEQ
The norm of $\mcurvH{a}{}$ 
gives the mean extrinsic curvature of $S$
\EQ 
\frac{1}{|H|} |\trexcurv(H)| =\sqrt{ 2|\trexcurv(u)\trexcurv(v)| } = |H| .
\endEQ

Now, the normal part of the Dirichlet symplectic vector is given by 
the normal mean curvature vector
$\Pperp{\rm D}{a}{} = \perpmcurvH{a}{}
= \v{a}{} \trexcurv(u) - \u{a}{} \trexcurv(v)$, 
which simplifies to 
\EQ
\Pperp{\rm D}{a}{} = 
-\trexcurv(u) (\der{u})^a 
-\left( \trexcurv(u) \psi_S^2 -\trexcurv(v) \right) (\der{v})^a 
-\trexcurv(u) \frac{2\der{\theta}R}{R^2} (\der{\theta})^a
-\trexcurv(u) \frac{2\der{\phi}R}{R^2 \sin^2\theta} (\der{\phi})^a . 
\endEQ
This vector gives the direction of zero expansion of $S$ 
in $(\Rnum^4,\flat{}{})$.

A preferred null frame for $T(S)^\perp$ consists of 
\EQs 
&& 
\fixu{a}{} = \frac{1}{\sqrt{2}|H|}( \perpmcurvH{a}{} + \mcurvH{a}{} )
= \sqrt{ \frac{\trexcurv(v)}{\trexcurv(u)} } (\der{v})^a , 
\label{ex1aHPfr+}\\
&&
\fixv{a}{} = \frac{1}{\sqrt{2}|H|}( \perpmcurvH{a}{} - \mcurvH{a}{} )
= \sqrt{ \frac{\trexcurv(u)}{\trexcurv(v)} } \left(
(\der{u})^a +\psi_S^2 (\der{v})^a 
+ \frac{2\der{\theta}R}{R^2} (\der{\theta})^a
+\frac{2\der{\phi}R}{R^2 \sin^2\theta} (\der{\phi})^a \right) , 
\label{ex1aHPfr-}
\endEQs
which depend only on $S$ and $\flat{ab}{}$ 
but not on the Minkowski frame $\adfr{\mu}{a}$. 
In the preferred frame \eqrefs{ex1aHPfr+}{ex1aHPfr-}, 
the tangential part of the Dirichlet symplectic vector is given by 
\EQ
\Ppar{\rm D}{a}{} = \metric{}{ac} \fixv{b}{} \covder{c} \fixu{}{b} 
= \metric{}{ac} \v{b}{} \covder{c} (du)_b
+\frac{1}{2} \metric{}{ac}\der{c} \ln(\trexcurv(u)/\trexcurv(v)) . 
\endEQ
This simplifies to 
\EQs
\Ppar{\rm D}{a}{} = 
&& 
\frac{1}{R^2} \left( 
(\der{\theta})^a + \der{\theta}R (\der{v})^a \right) 
\der{\theta} \ln(\trexcurv(u)/\trexcurv(v)) 
\nonumber\\&& 
+ \frac{1}{R^2 \sin^2\theta} \left( 
(\der{\phi})^a + \der{\phi}R (\der{v})^a \right) 
\der{\phi} \ln(R^2 \trexcurv(u)/\trexcurv(v)) . 
\endEQs
Therefore, since the dual vector 
$\Ppar{\rm D}{}{a}= \frac{1}{2}\covSder{a}\ln(R^2 \trexcurv(u)/\trexcurv(v))$ 
is a gradient on $S$, 
the normal curvature of $S$ is zero. 

The complete Dirichlet symplectic vector is
\EQ 
\P{a}{} =  \Pperp{\rm D}{a}{} + \Ppar{\rm D}{a}{} 
\endEQ 
which depends only on $S$ and $\flat{ab}{}$. 
In particular, it is independent of choice of the original null frame 
\eqrefs{ex1afrperpS}{ex1afrS} on Minkowski space 
and of the corresponding frame 
\eqrefs{ex1aTperpSfr+}{ex1aTperpSfr-} on $\TperpS$.
Geometrically, the dual vector $\Pperp{\rm D}{}{a}$ provides a preferred
normal direction for a family of hypersurfaces 
defined to cut the light cone at $S$, with vanishing normal curvature. 

Finally, the commutator of the null frame \eqref{ex1anullnormal}
yields the normal part of the Neumann symplectic vector
\EQs
\Pperp{\rm N}{a}{} =
\PrperpS [\v{}{},\u{}{}]\upindex{a} 
&&
= \PrperpS [ 
\der{u} + \frac{\psi^2}{2} \der{v} +\frac{\der{\theta}V}{r^2} \der{\theta} 
+\frac{\der{\phi}V}{r^2 \sin^2\theta} \der{\phi}, \der{v} ]\upindex{a}|_S
\nonumber\\&&
= \frac{\sqrt{2}}{r} \PrperpS\left( 
\frac{\psi^2}{2} (\der{v})^a +\frac{\der{\theta}V}{r^2} (\der{\theta})^a
+\frac{\der{\phi}V}{r^2 \sin^2\theta} (\der{\phi})^a \right)|_S
\nonumber\\&&
= -2\frac{\psi_S^2}{R} (\der{v})^a
\endEQs
through $\der{v}\psi^2= -\sqrt{2} \psi^2/r$.
The tangential part of the Neumann symplectic vector is simply
$\Ppar{\rm N}{a}{} =\Ppar{\rm D}{a}{}$. 
Hence, this yields the complete Neumann symplectic vector
\EQ
\P{a}{} =  \Pperp{\rm N}{a}{} + \Ppar{\rm N}{a}{} , 
\endEQ 
which depends only on the congruence of spacelike 2-surfaces
$u=\const$, $v-V(\theta,\phi)=\const$, 
lying on the light cones in Minkowski space. 

\subsubsection*{ Constant curvature 2-sphere }

In the special case of a constant curvature 2-sphere $S$, 
viewed as embedded either in a hyperplane 
$t=t_0=\const$, $r=R=r_0=\const$, 
or in a light cone, 
$u=u_0=(t_0-r_0)/\sqrt{2}$, $v=V=v_0=(t_0+r_0)/\sqrt{2}$, 
the mean curvature vector of $S$ is simply
$\mcurvH{a}{} = {2\over r_0} (\der{r})^a$,
and the complete Dirichlet symplectic vector reduces to 
\EQ
\P{a}{} =  \perpmcurvH{a}{} 
= {2\over r_0} (\der{t})^a , 
\endEQ
while the complete Neumann symplectic vector vanishes.

\subsection{Spherically Symmetric Spacetimes}

In a spherically symmetric spacetime $(\Rnum\times\Sigma,\g{ab}{})$,
\EQ\label{ex2}
\g{ab}{} =
- e^{2\psi} (dt)_a(dt)_b + e^{-2\nu} (dr)_a(dr)_b
+ r^2( (d\theta)_a (d\theta)_b + \sin^2\theta (d\phi)_a (d\phi)_b ) ,
\endEQ
where $\psi=\psi(t,r)$ and $\nu=\nu(t,r)$,
consider a spacelike 2-surface $S$
given by an isometry sphere $r=r_0=const$ and $t=t_0=const$.
The metric on $S$ is
\EQ
\metric{ab}{}=
r_0{}^2 (d\theta)_a (d\theta)_b + r_0{}^2 \sin^2\theta (d\phi)_a (d\phi)_b
\endEQ
and the area of $S$ is $A(S)=4\pi r_0{}^2$.
Fix an orthonormal frame adapted to $S$ by
\EQ
\adfr{0}{a} =  e^\psi (dt)_a , \quad
\adfr{1}{a} =  e^{-\nu} (dr)_a , \quad
\adfr{2}{a} =  r (d\theta)_a , \quad
\adfr{3}{a} =  r\sin\theta (d \phi)_a .
\label{ex2fr}
\endEQ
The Ricci rotation coefficients of the frame
\EQ
\frcon{\lambda}{\mu\nu}(\vartheta)
= \adfr{a}{\lambda} \adfr{b\nu}{} \covder{a} \adfr{\mu}{b}
= 2\adfr{a}{\lambda} \adfr{b[\mu}{} \der{[a} \adfr{\nu]}{b]}
- \adfr{b\mu}{} \adfr{c\nu}{} \der{[b} \adfr{}{c]\lambda}
\endEQ
have the following non-vanishing components:
\EQs
\frcon{0}{10} &=& -(\der{r} e^\psi) e^\nu e^{-\psi} ,\\
\frcon{1}{01} &=& -(\der{t} e^\nu) e^{-\nu} e^{-\psi} ,\\
\frcon{2}{12} &=& -{e^\nu \over r}
= \frcon{3}{13} ,\\
\frcon{3}{23} &=& - {\cos\theta \over r \sin\theta} .
\endEQs

The trace of the extrinsic curvatures of $(S,\metric{ab}{})$
with respect to
the frame on $\TperpS$
\EQ\label{ex2TperpSfr}
\t{}{a} =\adfr{0}{a}|_S ,\
\s{}{a}=\adfr{1}{a}|_S , 
\endEQ
are respectively
\EQ
\trexcurv(t) = \metric{}{ab} \covder{a} \t{}{b}
=-(\frcon{2}{02}+\frcon{3}{03})|_S
= 0 , 
\endEQ
and 
\EQ
\trexcurv(s) = \metric{}{ab} \covder{a} \s{}{b}
=-(\frcon{2}{12}+\frcon{3}{13})|_S
= {2 e^{\nu(t_0,r_0)} \over r_0} . 
\endEQ
A preferred direction in $T(S)^\perp$ is given by
the mean curvature vector
\EQ
\mcurvH{a}{} = \s{a}{} \trexcurv(s) - \t{a}{} \trexcurv(t)
={2 e^{2\nu(t_0,r_0)} \over r_0} (\der{r})^a , 
\endEQ
which is spacelike (outside of any horizon). 
This vector gives
the direction of the minimum absolute spacelike expansion of $S$.
Furthermore, the value of the expansion is given by
the trace extrinsic curvature of $S$ with respect to
the unit vector in the direction $\mcurvH{a}{}$,
\EQ
\frac{1}{\sqrt{H^2}} \trexcurv(H)
= |\trexcurv(s)|
= {2 e^{\nu(t_0,r_0)} \over r_0} , 
\endEQ
which is equal to the norm of $\mcurvH{a}{}$.

The normal part of the Dirichlet symplectic vector is
given by the normal mean curvature vector
\EQ
\Pperp{\rm D}{a}{} = \perpmcurvH{a}{}
= \t{a}{} \trexcurv(s) - \s{a}{} \trexcurv(t)
= 2 {e^{\nu(t_0,r_0)} e^{-\psi(t_0,r_0)} \over r_0} (\der{t})^a . 
\endEQ
Here $\Pperp{\rm D}{a}{}$ is timelike (outside of any horizon), 
orthogonal to $\mcurvH{a}{}$,
with the same absolute norm as $\mcurvH{a}{}$.
Most significant,
$\Pperp{\rm D}{a}{}$ gives the direction of zero expansion of $S$.

A preferred orthonormal frame for $T(S)^\perp$ is
\EQ\label{ex2HPfr}
\fixt{a}{}
=\frac{1}{\sqrt{H^2}} \perpmcurvH{a}{}
=e^{-\psi} (\der{t})^a , \quad
\fixs{a}{}
=\frac{1}{\sqrt{H^2}} \mcurvH{a}{}
=e^{\nu}(\der{r})^a , 
\endEQ
which depend only on $S$ and $\flat{ab}{}$
but not on the chosen frame $\adfr{\mu}{a}$.
In the preferred frame \eqref{ex2HPfr}
the tangential part of the Dirichlet symplectic vector is
\EQ
\Ppar{\rm D}{a}{}=\metric{}{ac} \fixt{b}{} \covder{c} \fixs{}{b}
= (\adfr{2a}{} \frcon{2}{10} + \adfr{3a}{} \frcon{3}{10})|_S
= 0 , 
\endEQ
and thus the normal curvature of $S$ is zero.

Hence the complete Dirichlet symplectic vector is
\EQ
\P{a}{} = \Pperp{\rm D}{a}{} + \Ppar{\rm D}{a}{}
=2 {e^{\nu(t_0,r_0)-\psi(t_0,r_0)} \over r_0} (\der{t})^a , 
\endEQ
which depends only on $S$ and $\flat{ab}{}$.
In particular, it is independent of choice of
the original orthonormal frame \eqref{ex2fr} for $\g{ab}{}$
and \eqref{ex2TperpSfr} for $\perpmetric{ab}{}$.

To define the Neumann symplectic vector,
it is natural to use the \onfr/ \eqref{ex2HPfr}
extended off $S$ to the congruence of isometry spheres
$t=\const$, $r=\const$.
Then, for this extension,
the normal part of the Neumann symplectic vector
is given by the commutator
\EQ
\Pperp{\rm N}{a}{} =
\PrperpS [\fixt{}{},\fixs{}{}]\upindex{a}
= (\fixt{a}{} \frcon{0}{10} + \fixs{a}{} \frcon{1}{10})|_S
= \left( -(\der{r} e^\psi) e^\nu e^{-2\psi} (\der{t})^a
+  (\der{t} e^\nu) e^{-\psi} (\der{r})^a \right)|_S.
\endEQ
Since $\Ppar{\rm N}{a}{}=\Ppar{\rm D}{a}{}=0$,
the complete Neumann symplectic vector
with respect to the congruence of isometry spheres associated to $S$
is given by
\EQ
\P{a}{} = \Pperp{\rm N}{a}{} + \Ppar{\rm N}{a}{}
= e^{\nu(t_0,r_0)-\psi(t_0,r_0)} \left(
-\der{r}\psi(t_0,r_0) (\der{t})^a
+\der{t}\nu(t_0,r_0) (\der{r})^a \right) .
\endEQ

Finally, as a special case,
consider the Reissner-Nordstr\"om black hole spacetime
obtained for
\EQ
e^\psi=e^\nu=\sqrt{1-2m/r+q^2/r^2}
\endEQ
where $m=\const$ and $q=\const$ are the black hole mass and charge;
$q=0$ yields the Schwarzschild black hole.
The mean curvature vector of an isometry sphere $S$,
$t=\const,r=\const$, outside of the horizon 
is given by
\EQ
\mcurvH{a}{}
={2 \over r} \left( 1-{2m\over r} +{q^2\over r^2} \right)  (\der{r})^a
\endEQ
which gives the direction of
the minimum absolute spacelike expansion of $S$.
Furthermore, the value of the expansion is given by
the norm of $\mcurvH{a}{}$,
\EQ
|\trexcurv(s)|
={2 \over r} \sqrt{1-{2m\over r}+{q^2\over r^2}} . 
\endEQ
The complete Dirichlet symplectic vector is given by
\EQ
\P{a}{}
= {2 \over r} (\der{t})^a
\endEQ
which depends only on $S$ and $\flat{ab}{}$.
Note that $\P{a}{}$ is timelike, 
orthogonal to $\mcurvH{a}{}$,
with the same absolute norm as $\mcurvH{a}{}$,
and it gives the direction of zero expansion of $S$.

With respect to the congruence of isometry spheres associated to $S$,
the complete Neumann symplectic vector is given by
\EQ
\P{a}{} =
-\left( {m\over r^2}-{q^2\over r^3} \right)
\left( 1-{2m\over r} +{q^2\over r^2} \right)^{-1} (\der{t})^a .
\endEQ
The curl of this vector yields the sectional curvature normal to $S$.

\subsection{Axisymmetric Spacetimes}

Now consider a stationary axisymmetric spacetime
$(\Rnum\times\Sigma,\g{ab}{})$,
\EQs
\g{ab}{} = &&
- e^{2\psi} (dt)_a(dt)_b + e^{-2\nu} (dr)_a(dr)_b
+ e^{-2\mu_1} (d\theta)_a (d\theta)_b
\nonumber\\&&
+ e^{-2\mu_2} ((d\phi)_a -w (dt)_a)((d\phi)_b -w (dt)_b) , 
\label{ex3}
\endEQs
where \cite{Chandra}
$w=w(r,\theta)$,
$\psi=\psi(r,\theta)$, $\nu=\nu(r,\theta)$,
$\mu_1=\mu_1(r,\theta)$ and $\mu_2=\mu_2(r,\theta)$. 
Let $S$ be a spacelike 2-surface given by $r=r_0=const$ and $t=t_0=const$,
which has the metric
\EQ
\metric{ab}{}=
e^{-2\mu_1(r_0,\theta)} (d\theta)_a (d\theta)_b
+ e^{-2\mu_2(r_0,\theta)} (d\phi)_a (d\phi)_b . 
\endEQ
The area of $S$ is
$A(S)=2\pi\int_0^\pi e^{-\mu_1(r_0,\theta)-\mu_2(r_0,\theta)} d\theta$.
A natural orthonormal frame adapted to $S$ is given by
\EQ
\adfr{0}{a} =  e^\psi (dt)_a , \quad
\adfr{1}{a} =  e^{-\nu} (dr)_a , \quad
\adfr{2}{a} =  e^{-\mu_1} (d\theta)_a , \quad
\adfr{3}{a} =  e^{-\mu_2} \left( (d \phi)_a - w (d t)_a  \right) . 
\label{ex3fr}
\endEQ
The Ricci rotation coefficients of the frame
\EQ
\frcon{\lambda}{\mu\nu}(\vartheta)
= \adfr{a}{\lambda} \adfr{b\nu}{} \covder{a} \adfr{\mu}{b}
= 2\adfr{a}{\lambda} \adfr{b[\mu}{} \der{[a} \adfr{\nu]}{b]}
- \adfr{b\mu}{} \adfr{c\nu}{} \der{[b} \adfr{}{c]\lambda}
\endEQ
have the following non-vanishing components:
\EQs
&&
\frcon{0}{01} = e^{-\psi} e^{\nu} \der{r}e^\psi ,
\frcon{3}{01} = \frac{1}{2} e^{-\psi} e^{\nu} e^{-\mu_2}\der{r}w ,
\\
&&
\frcon{0}{02} = e^{-\psi} e^{\mu_1} \der{\theta}e^\psi ,
\frcon{3}{02} = \frac{1}{2} e^{-\psi} e^{\mu_1} e^{-\mu_2} \der{\theta} w ,
\\
&&
\frcon{1}{03} = \frac{1}{2} e^{-\psi} e^{\nu} e^{-\mu_2} \der{r}w ,
\frcon{2}{03} = \frac{1}{2} e^{-\psi} e^{\mu_1} e^{-\mu_2}\der{\theta}w ,
\\
&&
\frcon{1}{12} = e^{\nu} e^{\mu_1} \der{\theta}e^{-\nu} ,
\frcon{2}{12} = -e^{\nu} e^{\mu_1} \der{r}e^{-\mu_1} ,
\\
&&
\frcon{0}{13} = \frac{1}{2} e^{-\psi} e^{\nu} e^{-\mu_2} \der{r}w ,
\frcon{3}{13} = -e^{\nu} e^{\mu_2} \der{r}e^{-\mu_2} ,
\\
&&
\frcon{0}{23} = \frac{1}{2} e^{-\psi} e^{\mu_1} e^{-\mu_2} \der{\theta}w ,
\frcon{3}{23} =  -e^{\mu_1} e^{\mu_2} \der{\theta}e^{-\mu_2} .
\endEQs
The trace of the extrinsic curvatures of $(S,\metric{ab}{})$
with respect to
the frame on $\TperpS$
\EQ\label{ex3TperpSfr}
\t{}{a} =\adfr{0}{a}|_S ,\ 
\s{}{a}=\adfr{1}{a}|_S
\endEQ
are respectively
\EQ
\trexcurv(t) = \metric{}{ab} \covder{a} \t{}{b}
=-(\frcon{2}{02}+\frcon{3}{03})|_S
= 0 , 
\endEQ
and 
\EQ
\trexcurv(s) = \metric{}{ab} \covder{a} \s{}{b}
=-(\frcon{2}{12}+\frcon{3}{13})|_S
= -e^{\nu(r_0,\theta)} \der{r}\mu(r_0,\theta) 
\endEQ
where $\mu(r,\theta)= \mu_1+\mu_2$. 
A preferred direction in $T(S)^\perp$ is given by
the mean curvature vector
\EQ
\mcurvH{a}{} = \s{a}{} \trexcurv(s) - \t{a}{} \trexcurv(t)
=-e^{2\nu(r_0,\theta)} \der{r}\mu(r_0,\theta) (\der{r})^a , 
\endEQ
which is spacelike (outside of any horizon). 
This vector gives
the direction of the minimum absolute spacelike expansion of $S$.
Furthermore, the value of the expansion is given by
the trace extrinsic curvature of $S$ with respect to
the unit vector in the direction $\mcurvH{a}{}$,
\EQ
\frac{1}{\sqrt{H^2}} \trexcurv(H)
= |\trexcurv(s)|
= e^{\nu(r_0,\theta)} |\der{r}\mu(r_0,\theta)| , 
\endEQ
which is equal to the norm of $\mcurvH{a}{}$.

The normal part of the Dirichlet symplectic vector is
given by the normal mean curvature vector
\EQ
\Pperp{\rm D}{a}{} = \perpmcurvH{a}{}
= \t{a}{} \trexcurv(s) - \s{a}{} \trexcurv(t)
= -e^{-\psi(r_0,\theta)} e^{\nu(r_0,\theta)} \der{r}\mu(r_0,\theta)
\left( (\der{t})^a +w  (\der{\phi})^a \right) . 
\endEQ
Here $\Pperp{\rm D}{a}{}$ is timelike (outside of any horizon), 
orthogonal to $\mcurvH{a}{}$,
with the same absolute norm as $\mcurvH{a}{}$.
Most significant,
$\Pperp{\rm D}{a}{}$ gives the direction of zero expansion of $S$.

A preferred orthonormal frame for $T(S)^\perp$ is
\EQ\label{ex3HPfr}
\fixt{a}{}
=\frac{1}{\sqrt{H^2}} \perpmcurvH{a}{}
= e^{-\psi} \left((\der{t})^a  +w  (\der{\phi})^a \right) , \quad
\fixs{a}{}
=\frac{1}{\sqrt{H^2}} \mcurvH{a}{}
= e^{\nu} (\der{r})^a , 
\endEQ
which depend only on $S$ and $\flat{ab}{}$
but not on the chosen frame $\adfr{\mu}{a}$.
In the preferred frame \eqref{ex3HPfr}
the tangential part of the Dirichlet symplectic vector is
\EQ
\Ppar{\rm D}{a}{}=\metric{}{ac} \fixt{b}{} \covder{c} \fixs{}{b}
= (\adfr{2a}{} \frcon{2}{10} + \adfr{3a}{} \frcon{3}{10})|_S
= \frac{1}{2} e^{-\psi(r_0,\theta)} e^{\nu(r_0,\theta)}
\der{r}w(r_0,\theta) (\der{\phi})^a .
\endEQ
The curl of this vector yields the normal curvature of $S$.

Hence the complete Dirichlet symplectic vector is
\EQs
\P{a}{} &&
= \Pperp{\rm D}{a}{} + \Ppar{\rm D}{a}{}
\nonumber\\&&
= e^{-\psi(r_0,\theta)+\nu(r_0,\theta)} \left(
-\der{r}\mu(r_0,\theta) \left( (\der{t})^a +w  (\der{\phi})^a \right)
+ \frac{1}{2}\der{r}w(r_0,\theta) (\der{\phi})^a \right) , 
\endEQs
which depends only on $S$ and $\flat{ab}{}$.
In particular, it is independent of choice of
the original orthonormal frame \eqref{ex3fr} for $\g{ab}{}$
and \eqref{ex3TperpSfr} for $\perpmetric{ab}{}$.

To define the Neumann symplectic vector,
it is natural to use the \onfr/ \eqref{ex3HPfr}
extended off $S$ to the congruence of 2-surfaces
$t=\const$, $r=\const$.
With respect to this extension,
the normal part of the Neumann symplectic vector
is given by the commutator
\EQ
\Pperp{\rm N}{a}{} =
\PrperpS [\fixt{}{},\fixs{}{}]\upindex{a}
= (\fixt{a}{} \frcon{0}{10} + \fixs{a}{} \frcon{1}{10})|_S
=-e^{-\psi(r_0,\theta)} e^{\nu(r_0,\theta)} \der{r}\psi(r_0,\theta)
\left( (\der{t})^a +w  (\der{\phi})^a \right) . 
\endEQ
Since $\Ppar{\rm N}{a}{}=\Ppar{\rm D}{a}{}$,
the complete Neumann symplectic vector
in this congruence of 2-surfaces associated to $S$
is given by
\EQs
\P{a}{} &&
= \Pperp{\rm N}{a}{} + \Ppar{\rm N}{a}{}
\nonumber\\&&
=e^{-\psi(r_0,\theta) + \nu(r_0,\theta)} \left(
-\der{r}\psi(r_0,\theta) \left( (\der{t})^a +w  (\der{\phi})^a \right)
+ \frac{1}{2}\der{r}w(r_0,\theta) (\der{\phi})^a \right) . 
\endEQs
The curl of this vector yields the sectional curvature normal to $S$.

As a special case, consider the Kerr black hole spacetime obtained
for \EQs e^\psi = {\sqrt\Delta \rho \over \Upsilon} , e^{-\nu} =
{\rho \over \sqrt\Delta} , e^{-\mu_1} = \rho , e^{-\mu_2} =
{\Upsilon \sin\theta \over \rho} , w = {2 a m r \over \Upsilon^2}
\endEQs
where
\EQ
\Upsilon^2= (r^2+a^2)^2 -a^2\Delta \sin^2\theta ,\
\Delta=r^2-2m r+a^2 ,\
\rho^2= r^2 +a^2\cos^2\theta . 
\endEQ
Here $m=\const$ and $a=\const$ are the black hole mass and angular
momentum; $a=0$ yields the Schwarzschild black hole. The mean
curvature vector of a 2-surface $S$, $t=\const,r=\const$, 
outside the horizon 
is given by
\EQ
\mcurvH{a}{} 
={\Delta\over \rho^2} {\der{r} \Upsilon \over \Upsilon}  (\der{r})^a
= \frac{\Delta}{\rho^2\Upsilon^2} ( 2r(r^2+a^2) -a^2(r-m)\sin^2\theta )
(\der{r})^a
\endEQ
which gives the direction of
the minimum absolute spacelike expansion of $S$.
Furthermore, the value of the expansion is given by
the norm of $\mcurvH{a}{}$,
\EQ
|\trexcurv(s)|
= \frac{\sqrt{\Delta} }{ \rho\Upsilon^2 }
( 2r(r^2+a^2) -a^2(r-m)\sin^2\theta ) . 
\endEQ
The complete Dirichlet symplectic vector is given by
\EQs
\P{a}{} &&
= {\Upsilon\over \rho^2}
\left( {\der{r} \Upsilon\over \Upsilon}
\left( (\der{t})^a+w (\partial_\phi)^a \right)
+ {\der{r} w \over 2} (\partial_\phi)^a \right)
\nonumber\\&&
= \frac{ 2r(r^2+a^2)  -a^2(r-m)\sin^2\theta }{ \rho^2\Upsilon } (\der{t})^a
+ \frac{ am }{  \rho^2\Upsilon } (\der{\phi})^a
\endEQs
which depends only on $S$ and $\flat{ab}{}$.
Note that the normal part of $\P{a}{}$ is timelike, 
orthogonal to $\mcurvH{a}{}$,
with the same absolute norm as $\mcurvH{a}{}$,
and it gives the direction of zero expansion of $S$.

With respect to this congruence of 2-surfaces $S$,
the complete Neumann symplectic vector is given by
\EQs
\P{a}{} &&
= {\Upsilon\over \rho^2}
\left( \left( {\der{r} \Upsilon\over \Upsilon}
-{\der{r} \Delta \over 2\Delta}
-{\der{r} \rho \over 2\rho} \right)
\left( (\der{t})^a
+w (\partial_\phi)^a \right)
+ {\der{r} w \over 2} (\partial_\phi)^a \right)
\nonumber\\&&
= -\frac{ 2m }{ \rho^4\Upsilon\Delta }\left( 
(r^2+a^2)^2 (r^2-a^2) +a^2( (r^2+a^2)^2-4m r^3 )\sin^2\theta
\right) (\der{t})^a
\nonumber\\&&\quad
+\frac{ am }{ \rho^4\Upsilon\Delta }\left(
(r^2-a^2)^2 -4 r^3 (r-m) -a^2( a^2-r^2 )\sin^2\theta
\right) (\der{\phi})^a . 
\endEQs

\subsection{Homogeneous Isotropic Spacetimes}

Next, consider a Friedmann-Robertson-Walker spacetime
$(\Rnum\times\Sigma,\g{ab}{})$,
\EQ\label{ex4}
\g{ab}{} =
- (dt)_a(dt)_b +  a^2(t) \left( {1\over 1-kr^2 } (dr)_a(dr)_b
+ r^2( (d\theta)_a (d\theta)_b + \sin^2\theta (d\phi)_a (d\phi)_b ) \right) ,
\endEQ
where $k=0,1,-1$
($\Sigma$ is $\Rnum^3$ if $k=0,-1$ or $S^3$ if $k=1$)
corresponding to a spatially flat, spherical, or hyperbolic geometry
on $\Sigma$.
Let $S$ be a spacelike 2-surface
given by an isometry sphere $r=r_0=const$ and $t=t_0=const$.
The metric on $S$ is
\EQ
\metric{ab}{}=
a^2(t_0) r_0{}^2 \left( (d\theta)_a (d\theta)_b
+ \sin^2\theta (d\phi)_a (d\phi)_b \right)
\endEQ
and the area of $S$ is $A(S)=4\pi a(t_0) r_0{}^2$.
Fix an orthonormal frame adapted to $S$ by
\EQ
\adfr{0}{a} =  (dt)_a , \quad
\adfr{1}{a} =  {a(t) \over \sqrt{1-k r^2}} (dr)_a , \quad
\adfr{2}{a} =  a(t) r (d\theta)_a , \quad
\adfr{3}{a} =  a(t) r\sin\theta (d \phi)_a .
\label{ex4fr}
\endEQ
The Ricci rotation coefficients of the frame
\EQ
\frcon{\lambda}{\mu\nu}(\vartheta)
= \adfr{a}{\lambda} \adfr{b\nu}{} \covder{a} \adfr{\mu}{b}
= 2\adfr{a}{\lambda} \adfr{b[\mu}{} \der{[a} \adfr{\nu]}{b]}
- \adfr{b\mu}{} \adfr{c\nu}{} \der{[b} \adfr{}{c]\lambda}
\endEQ
have the following non-vanishing components:
\EQs
\frcon{0}{01} &=& {\dot{a}(t) \over a(t)}
= \frcon{2}{02} =\frcon{3}{03} ,\\
\frcon{1}{12} &=& { \sqrt{1-2 k r} \over a(t) r}
= \frcon{3}{13} ,\\
\frcon{3}{23} &=& - {\cos\theta \over a(t) r \sin\theta} . 
\endEQs
(Here, an over-dot ``$\cdot$'' denotes a derivative with respect to $t$.)

The trace of the extrinsic curvatures of $(S,\metric{ab}{})$
with respect to
the frame on $\TperpS$
\EQ\label{ex4TperpSfr}
\t{}{a} =\adfr{0}{a}|_S ,\ 
\s{}{a}=\adfr{1}{a}|_S ,
\endEQ
are respectively
\EQ
\trexcurv(t) = \metric{}{ab} \covder{a} \t{}{b}
=-(\frcon{2}{02}+\frcon{3}{03})|_S
= {2 \dot{a}(t_0) \over a(t_0)}
\endEQ
and
\EQ
\trexcurv(s) = \metric{}{ab} \covder{a} \s{}{b}
=-(\frcon{2}{12}+\frcon{3}{13})|_S
= {2 \sqrt{1-k r_0{}^2} \over r_0 a(t_0)} . 
\endEQ
A preferred direction in $T(S)^\perp$ is given by
the mean curvature vector
\EQ
\mcurvH{a}{} = \s{a}{} \trexcurv(s) - \t{a}{} \trexcurv(t)
= 2 \left( {1-kr_0{}^2 \over a(t_0)^2 r_0} (\der{r})^a
- {\dot{a}(t_0)\over a(t_0)} (\der{t})^a \right) . 
\endEQ
This vector gives
the direction of the minimum absolute spacelike expansion of $S$.
Furthermore, the value of the expansion is given by
the trace extrinsic curvature of $S$ with respect to
the unit vector in the direction $\mcurvH{a}{}$,
\EQ
\frac{1}{\sqrt{H^2}} \trexcurv(H)
= |\trexcurv(s)|
= {2\over a(t_0)}  \sqrt{{1-kr_0{}^2 \over r_0{}^2}-\dot{a}{}^2(t_0)} , 
\endEQ
which is equal to the norm of $\mcurvH{a}{}$.

The normal part of the Dirichlet symplectic vector is
given by the normal mean curvature vector
\EQ
\Pperp{\rm D}{a}{} = \perpmcurvH{a}{}
= \t{a}{} \trexcurv(s) - \s{a}{} \trexcurv(t)
= {2\sqrt{1-k r_0{}^2} \over a(t_0)} \left( {1\over r_0} (\der{t})^a
- {\dot{a}(t_0) \over a(t_0)} (\der{r})^a \right) . 
\endEQ
Here $\Pperp{\rm D}{a}{}$ is orthogonal to $\mcurvH{a}{}$,
with the same absolute norm as $\mcurvH{a}{}$.
Most significant,
$\Pperp{\rm D}{a}{}$ gives the direction of zero expansion of $S$.
Note that $\Pperp{\rm D}{a}{}$ is timelike (and $\mcurvH{a}{}$ is spacelike)
if and only if the acceleration of $\Sigma$ satisfies
$|\dot{a}(t_0)|\leq \sqrt{1-k r_0{}^2} /r_0$,
depending on the radius of $S$.

A preferred orthonormal frame for $T(S)^\perp$ is
\EQ\label{ex4HPfr}
\fixt{a}{}
=\frac{1}{\sqrt{H^2}} \perpmcurvH{a}{}
=(\der{t})^a, \quad
\fixs{a}{}
=\frac{1}{\sqrt{H^2}} \mcurvH{a}{}
= {\sqrt{1-kr^2} \over a(t)} (\der{r})^a , 
\endEQ
which depend only on $S$ and $\flat{ab}{}$
but not on the chosen frame $\adfr{\mu}{a}$.
In the preferred frame \eqref{ex2HPfr}
the tangential part of the Dirichlet symplectic vector is
\EQ
\Ppar{\rm D}{a}{}=\metric{}{ac} \fixt{b}{} \covder{c} \fixs{}{b}
= (\adfr{2a}{} \frcon{2}{10} + \adfr{3a}{} \frcon{3}{10})|_S
= 0 , 
\endEQ
and thus the normal curvature of $S$ is zero.

Hence the complete Dirichlet symplectic vector is
\EQ
\P{a}{} = \Pperp{\rm D}{a}{} + \Ppar{\rm D}{a}{}
={2\sqrt{1-k r_0{}^2} \over a(t_0)} \left( {1\over r_0} (\der{t})^a
- {\dot{a}(t_0) \over a(t_0)} (\der{r})^a \right) , 
\endEQ
which depends only on $S$ and $\flat{ab}{}$.
In particular, it is independent of choice of
the original orthonormal frame \eqref{ex4fr} for $\g{ab}{}$
and \eqref{ex4TperpSfr} for $\perpmetric{ab}{}$.

To define the Neumann symplectic vector,
it is natural to use the \onfr/ \eqref{ex4HPfr}
extended off $S$ to the congruence of isometry spheres
$t=\const$, $r=\const$.
Then, for this extension,
the normal part of the Neumann symplectic vector
is given by the commutator
\EQ
\Pperp{\rm N}{a}{} =
\PrperpS [\fixt{}{},\fixs{}{}]\upindex{a}
= (\fixt{a}{} \frcon{0}{10} + \fixs{a}{} \frcon{1}{10})|_S
= - \left( {\sqrt{1-k r_0{}^2} \over a(t_0)^2} \right)
\dot{a}(t_0) (\der{r})^a .
\endEQ
Since $\Ppar{\rm N}{a}{}=\Ppar{\rm D}{a}{}=0$,
the complete Neumann symplectic vector
with respect to the congruence of isometry spheres associated to $S$
is given by
\EQ
\P{a}{} = \Pperp{\rm N}{a}{} + \Ppar{\rm N}{a}{}
= - \sqrt{1-k r_0{}^2}{\dot{a}(t_0) \over a(t_0)^2} (\der{r})^a .
\endEQ

For an isometry sphere $S$, $r=\const,t=\const$,
in a time-symmetric hypersurface $\Sigma$,
since $\dot{a}(t)=0$
it follows that
$\mcurvH{a}{}$ is spacelike, $\Pperp{\rm D}{a}{}$ is timelike.
Then the complete Dirichlet symplectic vector is
\EQ
\P{a}{}= {2\sqrt{1-k r^2} \over r a(t)} (\der{t})^a , 
\endEQ
while the complete Neumann symplectic vector vanishes.

\subsection{Asymptotically flat spacetimes}

Consider an \af/ spacetime $(M,\g{ab}{})$
with $\g{ab}{}= \flat{ab}{} +\O(1/r)$
and $\der{c}\g{ab}{}= \O(1/r^2)$
as $r\rightarrow\infty$ at fixed $t$,
where $\flat{ab}{}$ is a flat metric \eqref{ex1}
in Minkowski spherical coordinates $t,r,\theta,\phi$.
Suppose the total ADM mass, $m$, of the spacetime $(M,\g{ab}{})$
is finite and positive. 
Then the metric has the asymptotic form 
\cite{ADM,Regge-Teitelboim}
\EQs
\g{ab}{} = &&
-( 1-2m/r+\O(1/r^2) ) (dt)_a(dt)_b + ( 1+2m/r+\O(1/r^2) ) (dr)_a(dr)_b
\nonumber\\&&
+ r^2( (d\theta)_a (d\theta)_b + \sin^2\theta (d\phi)_a (d\phi)_b 
+\O(1/r^3) )
\quad\eqtext{ as $r\rightarrow\infty$ at fixed $t$. }
\label{afex}
\endEQs
(Note that any gravitational radiation terms vanish in this limit.)
We first discuss the ADM energy-momentum vector 
\cite{Wald-book}.
In the spacetime $(M,\g{ab}{})$,
spatial infinity, $\spi$, can be represented
as the set of asymptotic 2-spheres $S_\infty$
given by $t=\const,r=\rightarrow\infty$,
which are regarded as being identified under asymptotic time translations
generated by the Killing vector $(\der{t})^a$ of $\flat{ab}{}$.
Now, for a spacelike hypersurface $\Sigma_t$, $t=\const$,
the ADM energy and momentum
in standard asymptotic Minkowski coordinates $\x{\mu}{}$ on $M$
are given by 
\cite{ChoquetBruhat}
\EQs
\P{}{\mu} = \cases{ \displaystyle
\sum_{\nu,\rho\neq 0} \frac{1}{ 16\pi} \int_{S_\infty}
\s{}{\nu} ( \der{\rho}\g{\rho\nu}{} - \der{\nu}\g{\rho\rho}{} ) dS
= m
&, $\mu=0$\cr \displaystyle
\sum_{\nu\neq 0} \frac{1}{ 16\pi} \int_{S_\infty}
( \s{}{\nu} \der{t}\g{\mu\nu}{} - \s{}{\mu} \der{t}\g{\nu\nu}{} ) dS
=0
&, $\mu=1,2,3$. \cr}
\endEQs
Hence, the ADM energy-momentum vector at spatial infinity
is represented by $(\P{\rm ADM}{})_a = m (dt)_a|_{S_\infty}$.
Let $n^a=(\der{t})^a |_{S_\infty}$. 
Then, geometrically, the vector
\EQ\label{ADMvec}
\frac{1}{m} (\P{\rm ADM}{})^a = - n^a
\endEQ
corresponds to an asymptotic stationary unit-norm Killing vector 
of the \af/ metric \eqref{afex}.

To proceed, 
let $S$ be any family of spacelike 2-surfaces 
that approaches the 2-sphere $S_\infty$
as $r\rightarrow\infty$ at fixed $t$.
Since the spacetime metric is asymptotic to the Schwarzschild metric,
we may use the results obtained in Example (B)
to calculate the Dirichlet and Neumann symplectic vectors 
in this limit. 

The Dirichlet symplectic vector is given by
\EQ\label{afPD}
(\P{\rm D}{})^a={2\over r}( 1-m/r ) t^a +\O(1/r^3) 
\endEQ
where $t^a=( 1 +m/r+\O(1/r^2) )(\der{t})^a$ 
is a unit timelike vector of $(M,\g{ab}{})$. 
For a comparison with $(\P{\rm ADM}{})^a$,
we scale $(\P{\rm D}{})^a$ by the area of $S$, $A(S)=4\pi r^2+O(1/r)$,
which yields
\EQ\label{scaledPD}
A(S) (\P{\rm D}{})^a=8\pi( r-m +\O(1/r) ) t^a . 
\endEQ
Note that the first term in this expression is singular
as $r\rightarrow\infty$.
It corresponds to the Dirichlet symplectic vector
for $S_\infty$ with respect to the flat metric $\flat{ab}{}$ on $M$.
We extend this vector in a natural geometrical manner 
from $S_\infty$ to $S$ by 
\EQ
(\P{\rm D}{\rm flat})^a= {2\over r} t^a
\endEQ
which depends only on the radius of $S$
and the timelike unit vector $t^a$ with respect to $\g{ab}{}$.
We now subtract $(\P{\rm D}{\rm flat})^a$ from $(\P{\rm D}{})^a$ to obtain
the normalized Dirichlet symplectic vector
\EQ\label{spiPD}
(\barP{\rm D}{})^a = A(S) ( (\P{\rm D}{})^a - (\P{\rm D}{\rm flat})^a )
= (-8\pi m +\O(1/r) ) t^a . 
\endEQ
Then the limit $r\rightarrow\infty$ yields a well-defined (finite) vector
associated to $S_\infty$ in terms of $t^a\rightarrow n^a$.
This establishes our main result.

\Proclaim{ Theorem 4.1. }{
For an \af/ spacetime $(M,\g{ab}{})$,
at spatial infinity
the normalized Dirichlet symplectic vector \eqref{spiPD}
is equal to $8\pi$ times the ADM energy-momentum vector \eqref{ADMvec},
\EQ
\frac{1}{8\pi} (\barP{\rm D}{})^a |_{S_\infty} = (\P{\rm ADM}{})^a = -m\, n^a
\endEQ
}
(The $8\pi$ factor reflects the normalization chosen for 
the Hamiltonian variational principle 
for the Einstein equations in \secref{matter}.)

We remark that the ADM vector $(\P{\rm ADM}{})^a$ can be derived
\cite{Wald-Iyer1}
directly from the symplectic structure of the Einstein equations
similarly to the analysis given in Sec.~3 in \Ref{Anco-TungI}
by using \af/ \bdc/s at $S_\infty$
in place of the Dirichlet \bdc/ at $S$
on the spacetime metric.

Finally, we discuss the Neumann symplectic vector.
Note that the normalized symplectic vector \eqref{spiPD}
is obtained from the locally constructed Dirichlet symplectic vector
$(\P{\rm D}{})^a$ associated to a spacelike 2-surface $S$,
where $(\P{\rm D}{})^a$ depends only on $S$ and $\g{ab}{}$.
In contrast,
the Neumann symplectic vector $(\P{\rm N}{})^a$ associated to $S$
also depends on a choice of congruence of 2-surfaces $S'$
diffeomorphic to $S$.
If a suitably parameterized null geodesic congruence through $S$
is used to define $(\P{\rm N}{})^a$,
it follows from Proposition~3.14 that
the normal part of $(\P{\rm N}{})^a$ vanishes.
Moreover, the tangential part of $(\P{\rm N}{})^a$
is equal to the tangential part of $(\P{\rm D}{})^a$.
Thus for any topological 2-sphere $S$ 
that approaches $S_\infty$ as $r\rightarrow\infty$,
by \Eqref{afPD}
the resulting vector $(\P{\rm N}{})^a$
is at most $\O(1/r^3)$ and is tangential to $S$.
Consequently,
if we consider the normalized symplectic vector
\EQ
(\barP{\rm N}{})^a = A(S) ( (\P{\rm N}{})^a - (\P{\rm N}{\rm flat})^a )
\endEQ
defined analogously to $(\barP{\rm D}{})^a$,
then
\EQ\label{afPN}
(\barP{\rm N}{})^a |_{S_\infty} =0 . 
\endEQ
Note that for an \af/ metric \eqref{afex},
as $S$ approaches $S_\infty$,
all null geodesic congruences through $S$
approach future and past null infinity, $\scri$,
and thereby provide a natural congruence of spacelike 2-spheres
$S_\scri$ associated to the 2-sphere $S_\infty$
representing spatial infinity, $\spi$.
In particular, $S_\scri$ are related to $S_\infty$
by null geodesic asymptotic isometries of $\g{ab}{}$.
Hence, the normalized symplectic vector \eqref{afPN}
effectively depends only on $S$ and $\g{ab}{}$
(including its asymptotic structure),
similarly to the vector \eqref{spiPD}.

\section{ Concluding remarks }
\label{conclusion}

In this paper we have considered the covariant symplectic structure 
associated to the Einstein equations with matter sources. 
One main result is that we derive a covariant Hamiltonian
under Dirichlet and Neumann type \bdc/s for both 
the gravitational field and matter fields 
in any fixed spatially bounded region of spacetime $(M,\g{ab}{})$,
allowing the time-flow vector $\tfvec{a}$ to be timelike, spacelike, or null. 

The Dirichlet and Neumann Hamiltonians evaluated on solutions of 
the coupled gravitational and matter field equations 
reduce to a surface integral over the spatial boundary 2-surface, $S$. 
(In fact, this result is known to hold for any diffeomorphism covariant
spacetime field theory
\cite{Wald-Lee}.)
For each of the \bdc/s this surface integral has the form of 
$\int_S \tfvec{a} \P{}{a} dS$ 
where $\P{}{a}$ is a locally constructed dual vector field 
associated to the 2-surface $S$ and \bdc/s,
which we call the Dirichlet and Neumann symplectic vectors. 
Similar results are discussed in \Ref{Nester1,Nester2}.

Our principle result is to show that 
the purely gravitational part of the Dirichlet symplectic vector 
$(\P{\rm D}{})^a$ 
has very interesting geometrical properties
when decomposed into its normal and tangential parts,
$\Pperp{\rm D}{a}{}$ and $\Ppar{\rm D}{a}{}$, 
with respect to $S$. 
First, $\Pperp{\rm D}{a}{}$ depends only on 
the 2-surface $S$ and spacetime metric $\g{ab}{}$
and thus yields a geometrical vector field normal to $S$ in spacetime. 
This vector $\Pperp{\rm D}{a}{}$ is shown to be orthogonal to 
the mean curvature vector of $S$ and, most importantly, 
it gives the direction of zero expansion of $S$ in spacetime, 
\ie/ $\Lie{P_\perp} \vol{ab}{}(S)=0$ where 
$\vol{ab}{}(S)$ is the area volume form of $S$. 
Furthermore, the norm of the vector $\Pperp{\rm D}{a}{}$ is equal to 
the product of the expansions of $S$ with respect to 
ingoing and outgoing null geodesics, $\innframe{a}{}$ and $\outnframe{a}{}$
(and is independent of parameterization of the geodesics). 
This expression is obviously related to 
the condition for a spatial 2-surface $S$ 
to be trapped (or marginally trapped), namely, 
$\excurv{+}{}\excurv{-}{}$ is positive (or zero) on $S$,
where $\Lie{\theta^\pm}\vol{ab}{}(S) = \excurv{\pm}{} \vol{ab}{}(S)$. 
Consequently, $S$ is trapped (or marginally trapped) precisely when 
$\Pperp{\rm D}{a}{}$ is spacelike (or null) on $S$. 
If this notion is applied to the ingoing and outgoing null geodesics
at each point $p$ on $S$ 
(\ie/ the pair of null geodesics through $p$ is 
``trapped'' (or ``marginally trapped'') 
if $\excurv{+}{}\excurv{-}{}$ is positive (or zero) at $p$),
then, in this sense, 
$\Pperp{\rm D}{a}{}$ measures point-wise how close 
$S$ is to being a trapped surface. 

In contrast, $\Ppar{\rm D}{a}{}$ depends not only on 
the 2-surface $S$ and spacetime metric $\g{ab}{}$
but also on a choice of \onfr/ or null frame 
for the normal tangent space $\TperpS$ of $S$. 
Geometrically, $\Ppar{\rm D}{a}{}$ is shown to be 
a connection for the normal curvature of $S$ in spacetime
and consequently changes by a gradient 
under a boost of the frame. 
However, if the normal vector $\Pperp{\rm D}{a}{}$ is non-null, 
then $\Pperp{\rm D}{a}{}$ and the mean curvature vector of $S$
comprise a preferred frame for $\TperpS$ 
and hence there exists 
a corresponding preferred tangential vector $\Ppar{\rm D}{a}{}$
(evaluated in this frame). 
Thus, in this situation, the complete Dirichlet symplectic vector 
is a well-defined geometrical vector field 
depending only on $S$ and $\g{ab}{}$. 
We refer to this as the invariant Dirichlet symplectic vector
associated to $S$. 

Apart from its geometrical interest, 
the Dirichlet symplectic vector is also related to 
definitions of canonical energy, momentum, and angular momentum
given by the value of the Dirichlet Hamiltonian 
for solutions of the Einstein (and matter) equations. 
In particular, 
we have shown that in an asymptotically flat spacetime 
in the limit of $S$ approaching spatial infinity $S_\infty$, 
the Dirichlet symplectic vector reduces in a suitable sense 
to the ADM energy-momentum vector.
Hence, the integral 
$\int_{S_\infty} \tfvec{a} (\P{\rm D}{})_a dS$ 
yields total energy, momentum, angular momentum of the spacetime
when $\tfvec{a}$ is chosen to be 
an asymptotic Killing vector associated to 
time-translations, space-translations, or rotations
of the asymptotic flat background metric.

In addition, for a compact spatial 2-surface $S$ in $(M,\g{ab}{})$,
it follows from results in \Refs{Wald-Iyer2}
that the quasi-local quantities 
$\int_S \tfvec{a} (\P{\rm D}{})_a dS$ 
for $\tfvec{a}$ chosen to be normal and tangential to $S$
reproduce Brown and York's \cite{Brown-York1,Brown-York2}
quasilocal energy, momentum, and angular momentum quantities. 
(See also \Refs{Anco-TungI,Epp}.)
Furthermore, we have obtained matter contributions to these quantities,
for an electromagnetic field and a set of Yang-Mills-Higgs fields. 
In a forthcoming paper we will explore 
geometrical quasi-local quantities defined purely in terms of 
$\Pperp{\rm D}{a}{}$ and $\Ppar{\rm D}{a}{}$. 
We will also explore the use of $(\P{\rm D}{})^a$ 
as a time flow vector for a boundary-initial value formulation of 
the Einstein equations.

\acknowledgments
The authors thank the organizers of the 2nd workshop on Formal Geometry
and Mathematical Physics where this work was initiated. 
Bob Wald is thanked for helpful discussions.

\end{document}